\documentclass[twocolumn, appendixfloats, revtex4, numberedappendix, twocolappendix, iop]{openjournal}
\usepackage[pass]{geometry}

\usepackage{CJK}

\usepackage{enumitem}
\usepackage{url}

\usepackage[most,breakable]{tcolorbox}

\usepackage{orcidlink}

\begin{document}
\begin{CJK*}{UTF8}{gbsn}

\title{Scaling Laws for Emulation of Stellar Spectra \vspace{-1.2cm}}

\author{Tomasz R\'o\.za\'nski$^1$ \orcidlink{0000-0002-5819-3023}}
\email{Tomasz.Rozanski1@anu.edu.au}
\affiliation{Research School of Astronomy \& Astrophysics, The Australian National University, Cotter Rd., Weston, ACT 2611, Australia}
\affiliation{Astronomical Institute, University of Wroc\l aw, Kopernika 11, 51-622 Wroc\l aw, Poland}

\author{Yuan-Sen Ting (丁源森) \orcidlink{0000-0001-5082-9536}}
\affiliation{Department of Astronomy, The Ohio State University, Columbus, OH 43210, USA}
\affiliation{Center for Cosmology and AstroParticle Physics (CCAPP), The Ohio State University, Columbus, OH 43210, USA}

\begin{abstract}
Neural network-based emulators for the inference of stellar parameters and elemental abundances represent an increasingly popular methodology in modern spectroscopic surveys. However, these approaches are often constrained by their emulation precision and domain transfer capabilities. Greater generalizability has previously been achieved only with significantly larger model architectures, as demonstrated by Transformer-based models in natural language processing. This observation aligns with neural scaling laws, where model performance predictably improves with increased model size, computational resources allocated to model training, and training data volume. In this study, we demonstrate that these scaling laws also apply to Transformer-based spectral emulators in astronomy. Building upon our previous work with TransformerPayne and incorporating Maximum Update Parametrization techniques from natural language models, we provide training guidelines for scaling models to achieve optimal performance. Our results show that within the explored parameter space, clear scaling relationships emerge. These findings suggest that optimal computational resource allocation requires balanced scaling. Specifically, given a tenfold increase in training compute, achieving an optimal seven-fold reduction in mean squared error necessitates an approximately 2.5-fold increase in dataset size and a 3.8-fold increase in model size. This study establishes a foundation for developing spectral foundational models with enhanced domain transfer capabilities.
\end{abstract}


\maketitle

\section{Introduction}
\label{sec:intro}

The study of stellar spectra and the extraction of stellar properties is a critical component of astronomy, recently advanced by numerous spectroscopic surveys including APOGEE, LAMOST, Gaia-ESO, and GALAH \citep{Gilmore2012,Luo2015,Majewski2017,Buder2020}, with further progress anticipated from upcoming surveys such as 4MOST and WEAVE. The 4MOST survey \citep{2019Msngr.175....3D}, for instance, plans to collect approximately 20 million spectra at low resolution (R $\approx$ 6500) and 3 million at medium resolution (R $\approx$ 20,000) within a five-year timeframe. However, the unprecedented scale of these data necessitates more sophisticated automated pipelines for parameter inference to fully realize the potential of these surveys.

Machine learning models for spectral emulation have gained widespread adoption in this context. Since ab-initio spectrum synthesis, even under one-dimensional local thermodynamic equilibrium (1D-LTE) assumptions, remains computationally intensive, performing full spectral fitting with real-time generation can be impractical for large surveys. Emulators help amortize computational costs by learning the mapping between stellar parameters and their corresponding spectra, allowing for rapid inference once trained. As stellar spectra depend on dozens of parameters (stellar atmospheric parameters and chemical abundances across much of the periodic table), emulators must effectively handle high-dimensional interpolation, circumventing the ``curse of dimensionality'' that plagues traditional grid-based interpolation methods. 

Several approaches have been developed to address these needs, including quadratic models such as \textit{The Cannon} \citep{2015ApJ...808...16N}, a polynomial-based ridge regression; neural network-based models such as \textit{The Payne} \citep{2019ApJ...879...69T}; and \textit{TransformerPayne} \citep{2024arXiv240705751R}.

The success of emulators often stems from two key considerations: (a) how accurately the models can emulate synthetic spectra, that is, for a given set of fixed stellar properties, how closely the emulator-generated spectrum matches the ab-initio calculation; and (b) how effectively models trained on one domain of synthetic spectra can be transferred to other domains (whether between different theoretical frameworks, e.g., 1D LTE versus 3D non-LTE, or from synthetic spectra to empirical observations).

These considerations have driven exploration of various architectures. Simpler models such as ridge regression and fully connected networks often suffer from lower emulation accuracy at a given training size, largely due to inadequate inductive bias, which is the set of assumptions that a learning algorithm uses to help to generalize its prediction for unseen data. In our previous study, we proposed harnessing Transformer-based models \cite{2024arXiv240705751R} because of their appropriate inductive bias for capturing long-range information in spectra. This capability is particularly valuable when spectral lines from the same atomic species can be widely separated in wavelength or pixel space. 

We demonstrated that such Transformer models have the advantage of continued performance improvement with larger training sets and number of training steps, without encountering the performance plateaus common to simpler architectures. The observation of an apparent continuous improvement law, where a loss metric improves as a neural network's model size (measured by the number of free parameters or the compute required for a single evaluation), training duration, and training dataset size all increase, led to what is now widely known in machine learning as \emph{scaling laws} \citep[see][and references therein]{2017arXiv171200409H}.

The idea of scaling laws in large language models has been thoroughly studied in \citet{2020arXiv200108361K, 2022arXiv220315556H}, demonstrating that when model size, training length, and dataset size are scaled together in a linear fashion on a log scale, the validation loss continues to improve predictably over many orders of magnitude. This finding inspired the development of increasingly large language models. The success of scaling laws has been a critical cornerstone in demonstrating that with larger models (and computational resources) one could continue to improve machine learning models, not only solving the emulation accuracy problem but also simultaneously allowing greater generalizability for domain transfer \citep{2020arXiv200514165B, 2024arXiv240217193Z} that have been the motivation of larger models.

The concept of scaling laws has also begun to be explored in astronomical research, including applications to stellar light curves \citep{2024arXiv240517156P} and galaxy images \citep{2024arXiv240402973W, 2024arXiv240514930S}. Demonstrating the validity of scaling laws represents a critical step toward developing deep learning foundational models for astronomy. Scale plays a crucial role, as the generalizability of such foundational models primarily emerges with increasing scale. Analogously, large language models trained predominantly on English text exhibit improved ability to adapt to new languages or specialized domains when their scale is increased, highlighting scale-dependent improvements in domain transfer capabilities.

While the aforementioned studies have established the existence of scaling laws in astronomical applications, the approach to training these models has often lacked systematic methodology. Training deep learning foundational models requires careful understanding of how to scale up architectures while maintaining optimal hyperparameters. Hyperparameter choices, such as learning rate, can dramatically influence model performance. Therefore, knowing how to appropriately adjust these parameters according to model size, without resorting to expensive hyperparameter searches, represents a key step toward developing any foundational model, a challenge we aim to address specifically for stellar spectra in this work.

In this paper, we investigate scaling laws for the emulation of stellar spectra. We examine scaling with respect to training dataset size, neural network size, and training duration, thereby establishing a foundation for spectroscopic foundational models. This paper proceeds as follows: Section \ref{sec:Methods} presents the TransformerPayne architecture, the training dataset, the Maximum Update Parameterization ($\mu$P), and neural scaling laws. Section \ref{sec:results} describes the conducted experiments, including hyperparameter optimization, validation of $\mu$P, optimization of TransformerPayne architectural parameters, and experiments establishing scaling laws. Results are discussed in Section \ref{sec:discussion}, and conclusions are provided in Section \ref{sec:conclusion}.

\section{Methods}
\label{sec:Methods}

The goal of this study is to understand how emulator quality evolves as we scale to larger training datasets, larger neural networks, and longer training durations. We formulate the emulation problem following our previous work on the TransformerPayne model \citep{2024arXiv240705751R}. 

In these models, we construct a function $f_\theta(\lambda, \mathbf{p})$ which approximates exact normalized flux, where $\lambda$ represents the wavelength and $\mathbf{p}$ denotes a collection of stellar properties. In this work, the parameter vector $\mathbf{p}$ consists of one hundred parameters, including effective temperature, surface gravity, and elemental abundances spanning from helium to einsteinium. This function is implemented as a neural network parameterized by free parameters $\theta$, which are optimized to make $f_\theta(\lambda, \mathbf{p})$ an accurate approximation of spectra across the domain of interest.

TransformerPayne is a neural network constructed using several computational blocks. Two types of embedding modules, Sinusoidal Embedding and Multilayer Perceptron (MLP) Embedding, process the inputs to embed them into higher-dimensional spaces. These embeddings are then consecutively processed by $N$ TransformerBlocks. The first TransformerBlock takes the normalized embedding of the spectrum parameters, $\mathbf{p_\text{emb}}$, together with the sinusoidal embedding, $\mathbf{w_\text{emb}}$, of the input wavelength. All remaining TransformerBlocks replace the input $\mathbf{w_\text{emb}}$ with the output from the previous TransformerBlock, denoted as $\mathbf{r_i}$. The output of the last TransformerBlock is processed by an MLP head, which predicts a normalized flux (a scalar value) based on the final $\mathbf{r_N}$. 

In contrast to the original paper, we incorporate a simplified residual connection scheme that adheres to the Pre-LN placement \citep{2020arXiv200204745X}. This modification does not significantly impact the model's quality while aligning it more closely with Transformer architecture variants explored in the study introducing Maximum Update Parametrization (for details, see Sec.\,\ref{sec:mu_parameterization}), a set of hyperparameter optimization approaches that have shown promise in scaling up large language models.

For a comprehensive overview, we refer readers to the original TransformerPayne paper. Nonetheless, we provide a concise outline of the architecture below.

\begin{center}
    \begin{figure}[!ht]
        \centering
    	\includegraphics[scale=0.84]{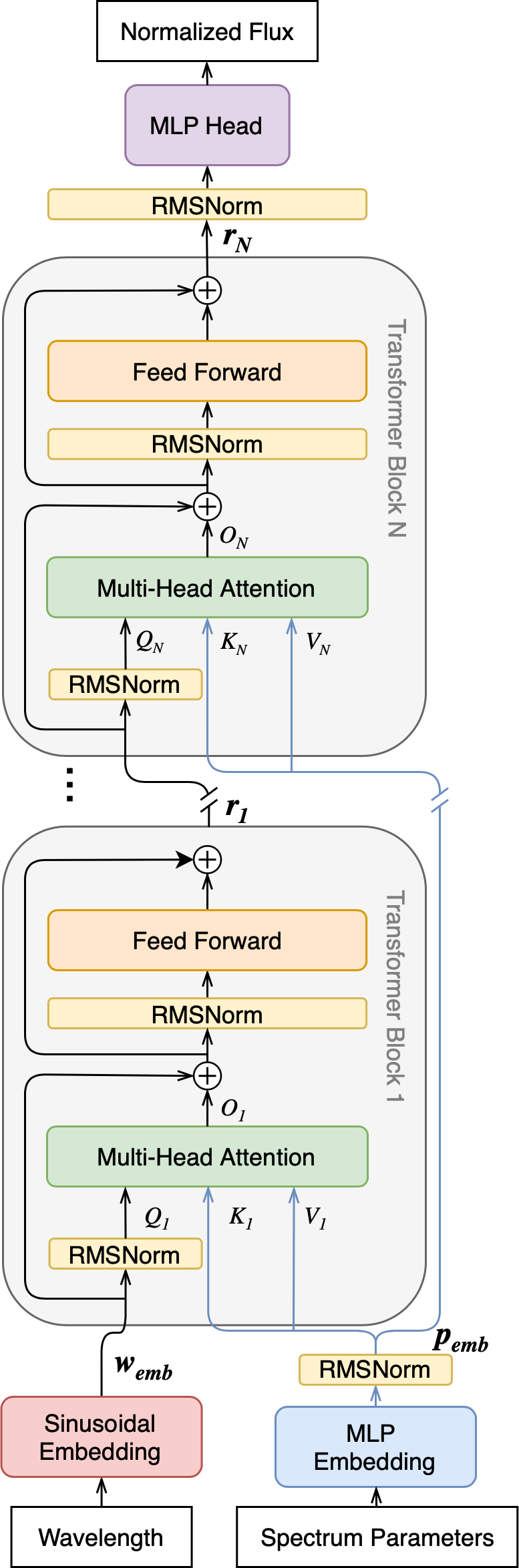}
    	\caption{Architecture diagram of the TransformerPayne variant implemented for our scaling experiments. The model processes two inputs: wavelength (lower left) and spectrum parameters (lower right). These inputs are transformed through Sinusoidal Embedding and MLP Embedding respectively, then normalized via RMSNorm. The embedded representations flow through N sequential Transformer blocks, each containing Feed Forward networks and Multi-Head Attention mechanisms with residual connections. Two Transformer blocks are shown for clarity; the ellipsis symbol and disconnections near $\mathbf{r_1}$ indicate that the remaining $N - 2$ blocks are omitted from the figure. The circled "+" symbols represent element-wise summation in residual connections. Each block processes outputs from the previous block and maintains the information flow through normalization layers. The final representation is processed by an MLP Head to predict the normalized flux. This architecture incorporates the Pre-LN placement scheme for residual connections, which facilitates stable training dynamics when scaling to larger model configurations.}
        \label{fig:fig1_Tpayne_scaling}
    \end{figure}
\end{center}

\subsection{MLP and Sinusoidal Embedding}

The inputs to TransformerPayne are $\lambda$, a scalar representing the wavelength, and $\mathbf{p}$, a vector of parameters. Since Transformer architectures were originally designed for language models that process discrete tokens (vector representations of words or subwords), adapting them to continuous astronomical data requires appropriate embedding strategies. For spectral modeling, there is no obvious natural tokenization, so we must transform both wavelength and stellar parameters into vector representations that Transformer models can effectively process and learn to correlate.

The wavelength is encoded using sinusoidal embedding given by the equation:
\begin{equation}
\label{eq:MLP_sin}
w_\text{emb,i} = \sin(\omega_i \lambda), \quad i = 1, \dots, d.
\end{equation}
where $\omega_i$ follow a geometric progression covering a manually adjusted range of circular frequencies, from $\omega_\text{min}$ to $\omega_\text{max}$, and index $i$ from 1 to $d$. We select these frequencies to cover all relevant scales, from continuum-level variations across the full 4000--5000\,\AA\ interval down to features smaller than the narrowest lines. Concretely, working in $\log_{10}\lambda$ (\,\AA), we take periods, $P$, geometrically spaced from 10 to $10^{-6}$ and convert them to angular frequencies via $\omega_i = 2\pi / \text{P}_i.$ The longest period tracks more than the entire spectral window, while the shortest resolves scales of order 0.05\,\AA ~and smaller - about a resolution of our spectral grid. This configuration performs well in practice, as confirmed by initial experiments, although fine-tuning the frequency range or spacing scheme could further improve emulation accuracy.

While the wavelength is embedded into a sequence consisting of a single vector, the spectral parameters vector is embedded into a sequence of $t$ tokens. This embedding is performed using an MLP Embedding, which is a simple two-layer perceptron. The output vector is then reshaped into a matrix with dimensions $t \times d$ and followed by RMSNorm layer for normalization. The function computed by the MLP Embedding is:
\begin{equation}
\label{eq:MLP_emb}
f(p_{\text{emb}}) = \alpha_\text{E} \mathbf{W}^{E}_{2}~\text{gelu}\big(\mathbf{W}^{E}_{1} \mathbf{p}\big),
\end{equation}
where $\mathbf{p}$ represents a vector of spectral parameters, $\mathbf{W}^{E}_{1}$ and $\mathbf{W}^{E}_{2}$ are weight matrices with shapes $d_{\text{p}} \times d$ and $d \times (t \cdot d)$ respectively, $\text{gelu}(x)$ is an element-wise non-linearity function, and $\alpha_E$ is a non-trainable, scalar hyperparameter.

To summarize, the embedding choices reflect domain-specific inductive biases tailored to the characteristics of the stellar spectra. Wavelengths are mapped using a fixed sinusoidal embedding to provide high-frequency basis functions that support the representation of sharp spectral features without increasing model complexity. By contrast, stellar parameters influence the flux more smoothly, so a lightweight two-layer MLP embedding is sufficient to capture their effects. While effective in practice, these choices are not necessarily optimal; more expressive embeddings may offer further gains and are a subject for future exploration.

\subsection{Transformer Block}

The core processing units of our architecture are the Transformer Blocks, which integrate information between wavelength representation and stellar parameter embeddings. Each Transformer Block consists of two main modules: Multi-Head Attention (MHA) and Feed-Forward (FF). Unlike the original TransformerPayne, our implementation employs a Pre-LN residual connection scheme \citep{2020arXiv200204745X}, which places the normalization before each sub-module rather than after, facilitating more stable training dynamics especially in deeper networks.

The Transformer Block is repeated $N$ times in sequence, with each block building upon the representations learned by the previous one. Each block receives three inputs: a query, a key, and a value. The key and value inputs are always the same collection of tokens encoding the parameters, $\mathbf{p_\text{emb}}$. The query input for the first block is the wavelength embedding, $\mathbf{w_\text{emb}}$, and for subsequent blocks, the query is the output of the previous block, $\mathbf{r_i}$. This recursive structure allows the model to progressively refine its understanding of the relationship between stellar parameters and spectral features. See Fig.\,\ref{fig:fig1_Tpayne_scaling} for a visual representation of this architecture.

Each input to the Attention block is treated as a sequence of tokens represented as a matrix in $\mathbb{R}^{(\text{sequence length}) \times d}$. In our case:
\begin{itemize}
  \item The wavelength embedding is a single-token sequence: $ \mathbf{w_\text{emb}} \in \mathbb{R}^{1\times d}$.
  \item The parameter tokens are $ \mathbf{p_\text{emb}} \in \mathbb{R}^{t\times d}$.
\end{itemize}
Hence, the key/value each have length $t$, while the query has length 1. This asymmetry reflects the core operation of our model: using the wavelength (query) to interrogate the stellar parameters (key/value) to determine the appropriate flux at that wavelength.

\paragraph{RMSNorm.}  
Before each sub-module (MHA or FF), we apply RMSNorm to normalize its input. Normalization is crucial for maintaining stable activations throughout the network and preventing internal covariate shift during training. 

We are not using any trainable scale or bias parameters, so the function computed by the RMSNorm of a vector $\mathbf{x} \in \mathbb{R}^{d}$ is simply:  
\begin{equation}
\label{eq:RMSNorm}
\text{RMSNorm}(\mathbf{x}) \;=\; \frac{\mathbf{x}}{\sqrt{\frac{1}{d}\,\sum_{i=1}^{d} x_i^2 + \epsilon}} \,,
\end{equation}  
where $\epsilon$ is a small constant (e.g., $10^{-6}$) for numerical stability. When applied to a sequence of tokens, like $\mathbf{p_\text{emb}} \in \mathbb{R}^{t\times d}$, this normalization is applied to every token individually, preserving the relative information within each token while standardizing the overall scale.

\paragraph{Multi-Head Attention.}
The Multi-Head Attention (MHA) mechanism is fundamental to the Transformer architecture, allowing the model to attend to different aspects of the stellar parameters simultaneously when generating spectral predictions. It serves as the primary mechanism for capturing relationships between wavelength positions and the various stellar parameters that influence the flux at those positions.

We first linearly project the RMS-normalized inputs into query, key, and value spaces. 
\begin{equation}
\label{eq:Q}
\mathbf{Q} \;=\; \text{RMSNorm}(\mathbf{x})\,W^{Q}, 
\end{equation}
\begin{equation}
\label{eq:K}
\mathbf{K} \;=\; \text{RMSNorm}(\mathbf{p_\text{emb}})\,W^{K},
\end{equation}
\begin{equation}
\label{eq:V}
\mathbf{V} \;=\; \text{RMSNorm}(\mathbf{p_\text{emb}})\,W^{V},
\end{equation}
where $\mathbf{x} \in \mathbb{R}^{1\times d}$ is either $\mathbf{w_\text{emb}}$ (for the first block) or $\mathbf{r_{i-1}}$ (for subsequent blocks). The matrices $W^Q, W^K, W^V \in \mathbb{R}^{d \times d}$. We split $\mathbf{Q},\mathbf{K},\mathbf{V}$ into $h$ heads, each of last dimension equal $d_\text{head} = d/h$. This multi-head approach allows different attention heads to specialize in detecting different types of spectral features or parameter interactions.

The dot-product attention for head $i$ is:
\begin{equation}
\label{eq:attention}
\mathbf{Z}_{i} \;=\; \text{softmax}\!\Bigl(
    \frac{\alpha_\text{att}\mathbf{Q}_i\mathbf{K}_i^\mathsf{T}}{d/h}
\Bigr)\,\mathbf{V}_i, 
\quad i = 1,\ldots,h,
\end{equation}
where shape of $\mathbf{Q}_i$ is $(1\times d_\text{head})$ and $\mathbf{K}_i$ shape is $(t\times d_\text{head})$, and $\alpha_\text{att}$ is a scalar non-trainable hyperparameter. Since there is only one query token, the attention map has a single row. 

This attention mechanism effectively allows the wavelength representation to selectively focus on relevant aspects of the stellar parameters that determine the flux at that wavelength. Then the head outputs $\mathbf{Z}_i \in \mathbb{R}^{1 \times d_\text{head}}$ are concatenated and linearly transformed:
\begin{equation}
\label{eq:O}
\mathbf{MHA}(\mathbf{x}, \mathbf{p_\text{emb}})
\;=\;
\bigl[\;\mathbf{Z}_{1},\dots,\mathbf{Z}_{h}\bigr]\,
W^{O},
\end{equation}
with $W^{O}\in \mathbb{R}^{d \times d}$.

\paragraph{Feed-Forward Network.}

After MHA, we again apply RMSNorm and then feed the result into a two-layer network. This Feed-Forward component provides additional non-linear processing capacity to each Transformer block, allowing it to model complex transformations beyond what attention alone can capture:
\begin{equation}
\label{eq:MLP_FF}
\text{FF}(\mathbf{u}) 
\;=\; 
W^{FF}_{2} \,\text{GELU}\!\bigl(W^{FF}_{1} \,\mathbf{u}\bigr),
\end{equation}
where $W^{FF}_{1}\in \mathbb{R}^{d \times d_{\text{ff}}}$ and $W^{FF}_{2}\in \mathbb{R}^{d_{\text{ff}} \times d}$. In all our experiments, we set $d_{\text{ff}}=4\,d$, which provides a sufficiently expanded intermediate representation for complex function approximation.

After the last Transformer block, the vector is processed with a two-layer network to predict the normalized flux. This final projection transforms the learned representation into the target output space:
\begin{equation}
\label{eq:MLP_head}
\text{normalized\,flux} 
\;=\; 
W^{H}_{2} \,\text{GELU}\!\bigl(W^{H}_{1} \,\text{RMSNorm}(\mathbf{r_N})\bigr),
\end{equation}
where $W^{H}_{1}\in \mathbb{R}^{d \times d}$ and $W^{H}_{2}\in \mathbb{R}^{1 \times d}$.

Each Transformer Block composes these sub-layers via Pre-LN residual connections, which help maintain gradient flow during training, especially in deeper networks:
\begin{equation}
\begin{aligned}
\mathbf{\hat{r}}_i \;&=\; \mathbf{x} \;+\; \mathbf{MHA}\!\Bigl(\text{RMSNorm}(\mathbf{x}),\, \text{RMSNorm}(\mathbf{p_\text{emb}})\Bigr),\\[6pt]
\mathbf{r}_i \;&=\; \mathbf{\hat{r}}_i \;+\; \text{FF}\!\bigl(\text{RMSNorm}(\mathbf{r}_i)\bigr),
\end{aligned}
\end{equation}
where $\mathbf{x} = \begin{cases}
\mathbf{w}_\text{emb}, & i=1\\
\mathbf{r}_{i-1}, & i>1
\end{cases}$.

We repeat this block $N$ times, creating a deep architecture capable of modeling the complex relationships between stellar parameters and spectral features across different wavelengths. The final single-token output $\mathbf{r}_N\in \mathbb{R}^{1\times d}$ is the transformed wavelength embedding, conditioned on the parameter tokens $\mathbf{p_\text{emb}}$, which contains all the information needed to predict the flux at the given wavelength.

\subsection{Dimensions of Scaling Laws}
\label{subsubsec:dim_of_scaling_law}

To systematically investigate how emulator performance improves with increased resources, we consider three dimensions of scaling: the size of the training data, the size of the neural network, and the computational resources used for training. In this section, we elaborate on how each of these dimensions is quantified and their relationships in our experimental framework.

First, we define the size of the training data as the number of distinct spectra, that is, spectra computed using different sets of atmospheric parameters. These training samples are drawn from a uniform distribution spanning the entire parameter space of interest, see Sec.\,\ref{sec:spectral_grid_training_metrics} for exact ranges. A uniformly sampled grid is straightforward to construct and is often sufficient for training; however, in regions of the parameter space that exhibit steep gradients, a locally denser or otherwise adaptive sampling strategy may be preferable. Such schemes may substantially improve sample efficiency and are left for future investigation.

The other two dimensions of scaling laws are the number of free parameters (neural network parameters optimized during training) and the total compute used for training. The number of free parameters directly reflects the model's capacity to represent complex relationships between stellar parameters and spectral features. It scales primarily with the model's width (embedding dimension $d$), depth (number of Transformer blocks $N$), and the number of tokens used to encode the parameters ($t$).

The computational cost, measured in floating-point operations (FLOPs), depends on these same architectural parameters, but also incorporates training-specific factors: the number of predicted wavelengths ($N_\text{flux}$), the batch size ($N_\text{batch}$), and the number of training steps ($S$). Together, these determine the total computational resources required to train a model to convergence. For additional details on parameter counts and FLOPs calculations for chosen modules, see Appendix\,\ref{appendix:params_and_flops}.

\begingroup
\setlength{\tabcolsep}{10pt}
\begin{table*}[ht!]
\centering
\caption{Summary of free parameters and floating-point operations  in Transformer Payne}
\begin{tabular}{l c c}
\hline
\textbf{Operation} & \textbf{Parameters} & \textbf{FLOPs} \\
\hline \hline

\multicolumn{3}{c}{\textit{Embedding Block}} \\
 \hline
\textbf{Sinusoidal Embedding, Eq.\,\ref{eq:MLP_sin}}
  & $0$
  & $(1 + \text{FLOP(sin)})\,N_{\text{flux}}\,d$
  \\

\textbf{MLP Embedding, Eq.\,\ref{eq:MLP_emb}}
  & $d_p\,d \;+\; t\,d^2$
  & $2\,d_p\,d \;+\; 2\,t\,d^2$
  \\

\textbf{RMSNorm, Eq.\,\ref{eq:RMSNorm}}
  & 0
  & $3\,t\,d$
  \\
 \hline
\multicolumn{3}{c}{\textit{Transformer Block} $\times N$} \\
 \hline
\textbf{Q/O Linear Layer, Eq.\,\ref{eq:Q}, \ref{eq:O}}
  & $N \,\times\, 2\,d^2$
  & $4\,N\,N_{\text{flux}}\,d^2$
  \\

\textbf{K/V Linear Layer, Eq.\,\ref{eq:K}, \ref{eq:V}}
  & $N \,\times\, 2\,d^2$
  & $4\,N\,t\,d^2$
  \\

\textbf{Multiplicative Attention, Eq.\,\ref{eq:attention}}
  & 0
  & $4\,N\,N_{\text{flux}}\,t\,d$
  \\

\textbf{RMSNorm, Eq.\,\ref{eq:Q}, \ref{eq:K}, \ref{eq:V}}
  & 0
  & $6\,N\,N_{\text{flux}}\,d$
  \\

\textbf{FeedForward, Eq.\,\ref{eq:MLP_FF}}
  & $N \,\times\, 8\,d^2$
  & $16\,N\,N_{\text{flux}}\,d^2$
  \\
 \hline
\multicolumn{3}{c}{\textit{Prediction Block}} \\
 \hline
\textbf{RMSNorm, Eq.\,\ref{eq:MLP_head}}
  & 0
  & $3\,N_{\text{flux}}\,d$
  \\

\textbf{MLP Head, Eq.\,\ref{eq:MLP_head}}
  & $d^2 \;+\; d$
  & $2\,N_{\text{flux}} \bigl(d^2 \;+\; d\bigr)$
  \\
\hline
\end{tabular}

\vspace{1em}
\parbox{\textwidth}{\small The FLOPs shown in the table are calculated for the vectorized version of TransformerPayne, which predicts a vector of normalized flux given a vector of wavelengths, $\mathbf{\lambda} \in \mathbb{R}^{N_\text{flux}}$. For the non-vectorized version of TransformerPayne, it is sufficient to substitute $N_\text{flux} = 1$. Usually, $N_\text{flux} > d > t$, and $d_p$ ranges from 5 to about 100. The leading terms in the FLOPs count are therefore those of order $d^2 \cdot N_\text{flux}$. The parameter count is dominated by the term quadratic in $d$. $\text{FLOP}(\sin)$ represents the cost of a sine function evaluation, which is typically on the negligible order of tens of FLOPs.}
\label{tab:tp_parameters_and_flops}
\end{table*}
\endgroup

A summary of the number of free parameters and FLOPs for each component of our architecture is provided in Table\,\ref{tab:tp_parameters_and_flops}, from which it follows that the exact number of free parameters, $P$, equals:
\begin{equation}
\text{P}
= (t + 12N + 1)d^{2} 
+ (d_{p} + 1)d.
\end{equation}

The leading terms of this equation are $(t + 12N)d^{2}$, so the hyperparameter primarily driving the number of free parameters is the dimensionality $d$. Here, $t$ represents the number of parameter tokens, $N$ is the number of Transformer blocks, and $d_p$ is the dimension of the input parameter vector. This quadratic relationship between model dimension and parameter count is characteristic of Transformer architectures and plays a crucial role in scaling behavior.

The approximate formula for the computational cost of the TransformerPayne forward pass (i.e., evaluation of neural network), as derived from Table~\ref{tab:tp_parameters_and_flops}, is:
\begin{equation}
\begin{aligned}
\text{Evaluation FLOPs} =\\
\bigl(2t + 20NN_{\text{flux}} + 4Nt + 2N_{\text{flux}}\bigr)d^{2}\\
+ \Bigl[(1 + \text{FLOP}(\sin)) + 5 + 6N\Bigr]N_{\text{flux}}d\\
+ \Bigl[3 + 4NN_{\text{flux}}\Bigr]td + 2d_{p}d.
\end{aligned}
\end{equation}

Here, $N_\text{flux}$ represents the number of wavelength points for which the flux is predicted simultaneously, which affects vectorization efficiency. We note that this is only an approximate expression, as it does not include the cost of evaluating activation functions or other minor costs due to multiplicative factors and residual connections. However, these costs are largely negligible compared to the leading terms. $\text{FLOP}(\sin)$ represents the cost of a single $\sin$ function evaluation, which is on the order of tens of FLOPs (for our calculations, we assumed $\text{FLOP}(\sin) = 10$).

To estimate the total computational cost of training, which is the key metric considered in scaling laws, the cost of a forward pass per single spectrum needs to be multiplied by a factor of 3. This factor accounts for the fact that each training step includes one forward pass and one backward pass for gradient evaluation, with gradient evaluation typically costing twice as much as the forward pass, a standard approximation in the literature \citep[e.g.][]{2020arXiv200108361K}. Additionally, this must be multiplied by the batch size and the number of training steps. The total training FLOPs, up to the leading terms, can be estimated using the expression:
\begin{equation}
\text{Total training FLOPs} \approx 3 S N_\text{batch} \times 20 N N_\text{flux} d^2.
\end{equation}

This formulation allows us to systematically explore how emulator performance scales with increasing model size, training data, and computational resources in our experimental investigations.

\subsection{Spectral Grid, Training and Metrics}
\label{sec:spectral_grid_training_metrics}

The experiments in this work utilize one of the synthetic datasets of 100,000 normalized stellar spectra introduced in \citet{2024arXiv240705751R}. These spectra were generated using plane-parallel Local Thermodynamic Equilibrium (LTE) atmospheric models \citep{2008AA...491..633L, 1979ApJS...40....1K, Kurucz2005}. The wavelength range spans from 4000 to 5000\,\AA~at a resolution of $R=100,000$, sampled at 22,315 wavelengths spaced equidistantly. The grid varies the effective temperature, $T_\text{eff}$, from 4000\,K to 6000\,K and the surface gravity from 4 to 5. The microturbulence velocity remains fixed at $\xi=0$\,km/s. Helium abundances vary between none and twice the solar value (assuming 0.0782 in the Sun), while the abundances of other elements (atomic numbers $Z=3$ to $Z=99$) are sampled uniformly between $-2$ and 1\,dex relative to the solar abundance. Although \citet{2024arXiv240705751R} showed that only 38 elemental abundances are well constrained on this spectral grid, we retain the full training dataset to ensure direct comparability with the results presented in that work. Adding parameters with no spectral effect does not compromise accuracy, as the model internally discards them through learned representations. It is important to note that these spectra are normalized using the theoretical continuum, resulting in flux values predominantly near 1 (with absorption features appearing as downward deviations below 1). Consequently, all MSE values reported in this study are with respect to this normalized scale, making them dimensionless quantities that directly represent fractional deviations in the normalized flux.

We adopt this dataset for convenience , but emphasize that our primary focus is on investigating the scaling properties of the TransformerPayne model. By scaling properties, we mean how model performance systematically improves as we increase model size, training data, and computational resources, relationships that would hold qualitatively even with different synthetic or empirical models, although the exact numerical values might differ.

We note that these spectra are normalized using the theoretical continuum, resulting in flux values predominantly near 1 (with absorption features appearing as downward deviations below 1). Consequently, all MSE values reported in this study are with respect to this normalized scale, making them dimensionless quantities that directly represent fractional deviations in the normalized flux.

We perform stochastic gradient-based optimization using the AdamW optimizer \citep{2014arXiv1412.6980K, 2017arXiv171105101L}, which incorporates adaptive learning rates, momentum-based updates, and decoupled weight decay regularization. All experiments were conducted with a batch size of $N_\text{batch} = 32$ spectra, each containing $N_\text{flux} = 1024$ flux values linearly interpolated at randomly sampled wavelengths. 

To control the global learning rate, denoted $\eta$, throughout training, we use the Warmup-Stable-Decay (WSD) scheduler (see, e.g., \citet{2024arXiv240518392H}, and references therein), which consists of three phases: (1) a linear warm-up from 0 to the maximum learning rate over the first $N_\text{warm-up}$ steps, (2) a stable phase where training continues at the maximum learning rate, and (3) a linear cool-down phase that decreases the learning rate to zero over $N_\text{cool-down}$ steps.

The WSD schedule is particularly useful for scaling law experiments, as it allows the warm-up phase to be fixed while deferring the cool-down phase until later in training. This flexibility significantly reduces the time required to test different training durations without incurring substantial computational overhead. In our initial tests, we verified that WSD performs at least as well as the linear warm-up followed by cosine decay schedule used in the TransformerPayne paper \citep{2024arXiv240705751R}, making it a suitable choice for our study. Additionally to global learning rate $\eta$, we also tuned an embedding's learning rate scaling factor $\eta_\text{E}$.

An important factor in training dynamics, alongside the learning rate, is the scale of weight initialization in the network. For all matrix initializations, we used a truncated Gaussian distribution (truncated at $2\sigma$) with a mean of $0$ and a standard deviation $\sigma$. Each matrix was initialized independently and proportionally to $1/\sqrt{d_\text{in}}$, with a globally tuned initialization scale $\sigma$.

As is standard practice for regression problems, we used the Mean Squared Error (MSE) as our loss function. It quantifies the average squared difference between the predicted and true normalized fluxes across all wavelengths and spectra:
\begin{equation}
\text{MSE} = \frac{1}{N} \sum_{i=1}^{N} \frac{1}{N_\text{flux}} \sum_{j=1}^{N_\text{flux}} \left( y_{ij} - f_\theta(\lambda_{ij}, \mathbf{p}_i) \right)^2,
\end{equation}
where $N$ represents the number of spectra (during training, $N = N_\text{batch}$, the batch size; during validation, $N = 1024$, the size of our validation set), $N_\text{flux}$ is the number of flux points per spectrum ($N_\text{flux} = 1024$), $\lambda_{ij}$ is the wavelength, $\mathbf{p}_i$ represents the stellar parameters, and $y_{ij}$ is the normalized flux for the $j$-th wavelength of the $i$-th spectrum. The function $f_\theta(\lambda_{ij}, \mathbf{p}_i)$ represents the neural network's prediction, parameterized by $\theta$. Note that MSE directly quantifies the squared flux error; therefore, if an estimate of the typical flux uncertainty (standard deviation) is desired, one should take the square root of MSE.

To evaluate model generalization and avoid overfitting, we maintained a validation dataset of 1024 spectra drawn from the same parameter domain as the training grid but not used during training. We computed the MSE on this validation dataset at predefined checkpoints throughout the training process. For consistent and fair comparisons across different experimental configurations, we report the lowest validation MSE recorded during the entire training process for each model. This approach ensures that we capture each model's optimal performance, regardless of when it occurs during training, while still measuring generalization to unseen data. Unless stated otherwise, all the MSE losses of the scaling law refer to the validation set, not the training set.

\subsection{Maximum Update Parametrization}
\label{sec:mu_parameterization}

As we move toward developing spectral foundational models with billions of parameters, the ability to train efficiently at scale becomes crucial. Conventional approaches like grid search for optimal training configurations become prohibitively expensive and wasteful at these scales. What is needed instead is a systematic, reliable training protocol that consistently produces high-quality models without extensive tuning, precisely the goal of our exploration in this study. 

Scaling the size of a neural network inherently involves two interconnected aspects. The first concerns the choice of architectural hyperparameter configurations for a fixed model size. This means determining how to allocate parameters when increasing model capacity, whether to add more Transformer blocks (increasing depth), expand the embedding dimension (increasing width), add more attention heads, or some combination of these approaches. In the case of TransformerPayne, these key hyperparameters, as mentioned in Sec.\,\ref{subsubsec:dim_of_scaling_law}, include the dimensionality, $d$; the number of Transformer blocks, $N$; the number of tokens used to encode the parameters, $t$; and the number of heads, $h$. 

The specific choice of these hyperparameters directly impacts the final performance of the model. In principle, one could scale by increasing only a single hyperparameter (e.g., only making the model deeper by adding more Transformer blocks while keeping width constant). However, this approach is suboptimal and does not yield realistic scaling laws. Therefore, a challenge is to understand how different hyperparameter choices affect model performance given a specific model size. A concrete example of this is determining how to balance depth ($N$) versus width ($d$) when scaling the model.

The second aspect involves tuning the hyperparameters related to the training process to ensure stable and efficient learning. Key parameters include the maximum learning rate, the scale of random weight initialization, and learning rates for specific network components. Improper tuning can lead to instability or ineffective weight updates, potentially masking improvements from scaling. This issue becomes particularly critical at larger model sizes, where exhaustive hyperparameter searches are computationally prohibitive due to the high cost of training multiple large-scale networks.

This challenge of optimizing hyperparameters efficiently at extreme scales underscores the need for systematic strategies that reduce the cost of hyperparameter tuning. A key development in this direction, originating in the context of large language models, has emerged from studies on the Maximum Update Parametrization ($\mu$P or $\mu$-Parametrization), which provides a principled framework \citep{2022arXiv220303466Y,2023arXiv231002244Y}. This theoretical framework demonstrates that when networks are scaled appropriately, certain hyperparameters remain stable across orders of magnitude of model sizes, enabling efficient transfer of tuning insights from smaller models to much larger ones.

The insight is that for meaningful evolution of hidden representations in neural networks, the spectral norm of the network weights ($\|\mathbf{W}\|_*$) and their updates ($\|\Delta\mathbf{W}\|_*$) should remain on the order of $\sqrt{m/n}$, where $\mathbf{W} \in \mathbb{R}^{m \times n}$ \citep{2023arXiv231017813Y}. This ensures that hidden features maintain their $l^2$-norms while their updates evolve proportionally to their magnitudes, preventing both the vanishing and exploding of signals through the network during training. When $\mathbf{A} \in \mathbb{R}^{m\times n}$, the spectral norm is defined by:
\begin{equation}
    \|\mathbf{A}\|_* := \max_{v \in \mathbb{R}^n \setminus \{0\}} \frac{\|\mathbf{A} v\|_2}{\|v\|_2}.
\end{equation}

While this theoretical perspective can be implemented through direct normalization of matrix norms during training, a more practical approach, as demonstrated in \citet{2023arXiv231017813Y}, involves appropriate initialization of weight scales for individual matrices in the neural network and corresponding adjustment of learning rates as the network's dimensionality changes. This is the approach developed in \citet{2022arXiv220303466Y} and the methodology we have implemented for the TransformerPayne scaling experiments presented in this work \citep[see also][for practical guidance]{2024arXiv240405728L,2024arXiv240717465B}.

To implement $\mu$P in our models, we systematically rescale the initialization scale and learning rate for individual matrices as the width of TransformerPayne, $d$, increases. When training neural networks with the Adam optimizer, the initialization scale $\sigma$ and learning rate $\eta$ for any matrix $\mathbf{W} \in \mathbb{R}^{m \times n}$ should follow these rescaling relationships:
\begin{equation} 
\sigma \propto \frac{1}{\sqrt{n}}\min \left[ 1,\sqrt{\frac{m}{n}}\right]
\end{equation}
\begin{equation}
\eta \propto \frac{1}{n},
\end{equation}
where $n$ corresponds to the dimension of the input to the linear layer $\mathbf{W}$, while $m$ represents the output dimension. In our implementation of TransformerPayne, we applied these appropriate scaling factors to all matrices throughout the architecture, ensuring consistent behavior across different model sizes.

\subsection{Neural Scaling Laws}
\label{sec:methods_neural_scaling_laws}

Neural scaling laws provide a framework for understanding how model performance systematically improves with increased resources. In the context of stellar spectra emulation, these laws characterize the relationship between validation loss and three critical scaling dimensions: model size ($P$), training dataset size ($D$), and computational resources used for training ($C$). By training TransformerPayne networks of various sizes on spectral grids of different sizes, we can empirically determine whether these relationships follow simple scaling laws, enabling us to forecast performance at scales beyond our current experiments.

Following the convention established by \citet{2020arXiv200108361K}, we consider three primary scaling regimes, each representing a scenario in which emulator performance is capped by a single scarce resource -- model parameters, training data, or training compute:

\begin{enumerate}
    \item \textbf{Parameter-limited regime}: When the number of free parameters is the limiting factor (assuming abundant data and compute):
    \begin{equation}
        \label{eq:scaling_model_size}
        \mathcal{L}(P) = \left(\frac{P_c}{P}\right)^{\alpha_P},
    \end{equation}
    
    \item \textbf{Data-limited regime}: When the size of the training dataset is the bottleneck (assuming sufficiently large models regularized using early stopping):
    \begin{equation}
        \label{eq:scaling_data_size}
        \mathcal{L}(D) = \left(\frac{D_c}{D}\right)^{\alpha_D},
    \end{equation}
    
    \item \textbf{Compute-limited regime}: When computational resources are the limiting factor (assuming a large dataset, an optimally-sized model, and a small fixed batch size):
    \begin{equation}
        \label{eq:scaling_compute}
        \mathcal{L}(C) = \left(\frac{C_c}{C}\right)^{\alpha_C},
    \end{equation}
\end{enumerate}

For the compute-limited regime, we note that our investigation is slightly constrained due to computational limitations. Since training with large batch sizes was not feasible in our experimental setup, we could not precisely investigate the critical batch size limit, the batch size at which training achieves an optimal balance between computational efficiency and optimization steps. For an in-depth discussion of critical batch size, we refer readers to \citet{2018arXiv181206162M}. This limitation, however, does not undermine the broader relevance of our findings.

In addition to analyzing the three primary scaling regimes, we extensively quantified the last one, the compute-limited regime, due to its practical significance. Specifically, we quantified how model size and the number of processed spectra scale along the frontier of compute-efficient models. We define a \textit{frontier model} as one that achieves the best MSE for a given training budget, effectively tracing a Pareto-optimal curve in the space of model size and training set size. A Pareto-optimal curve represents configurations where no improvement can be made in one dimension (e.g., model accuracy) without sacrificing performance in another dimension (e.g., computational cost). This approach enabled us to derive two additional power-law scaling relations describing how model size and the number of training spectra should grow when following the compute frontier, shedding further light on the interplay among data, model parameters, and total training cost.

\section{Results}
\label{sec:results}

Establishing robust scaling laws for stellar spectra emulation with TransformerPayne required careful optimization of hyperparameters and refinement of training strategies to prevent instabilities that could obscure the underlying empirical trends. Our investigation followed a systematic approach: first, we fine-tuned key training-related hyperparameters using smaller proxy models to determine optimal configurations efficiently. We then validated our $\mu$-Parametrization implementation to ensure stability across different scaling regimes. With these optimal training parameters in place, as we will demonstrate consistently across all settings, we proceeded to train a range of models varying in size, training duration, and training set size. This exploration allowed us to characterize the full scaling behavior of TransformerPayne across multiple dimensions of interest.

\subsection{Hyperparameter Optimization}

We began by optimizing training-related hyperparameters for TransformerPayne, specifically the maximum learning rate ($\eta$), the global initialization scale ($\sigma$), the embedding learning rate scale factor ($\eta_\text{E}$), and the embedding and attention scaling factors ($\alpha_{\text{E}}$ and $\alpha_{\text{att}}$). These parameters control different aspects of training: $\eta$ determines the step size during gradient descent, $\sigma$ affects the initial distribution of network weights, $\eta_\text{E}$ adjusts the relative learning rate for embedding parameters, while $\alpha_{\text{E}}$ and $\alpha_{\text{att}}$ scale the outputs of the embedding and attention layers, respectively.

To determine optimal values, we trained 100 small-scale TransformerPayne models (with $d=128$, $N=8$, $t=16$, and $h=4$) on a full training dataset consisting of 100,000 spectra. Each model was trained for 20,000 steps, including 2,000 warm-up steps followed by a cosine decay to zero. Hyperparameter values were sampled uniformly in log-space: the learning rate $\eta$ was drawn from $10^{-4}$ to $10^{-1}$, and all other hyperparameters were drawn from $0.1$ to $10$.
\begin{center}
    \begin{figure*}[!t]
        \centering
    	\includegraphics[width=1.0\hsize]{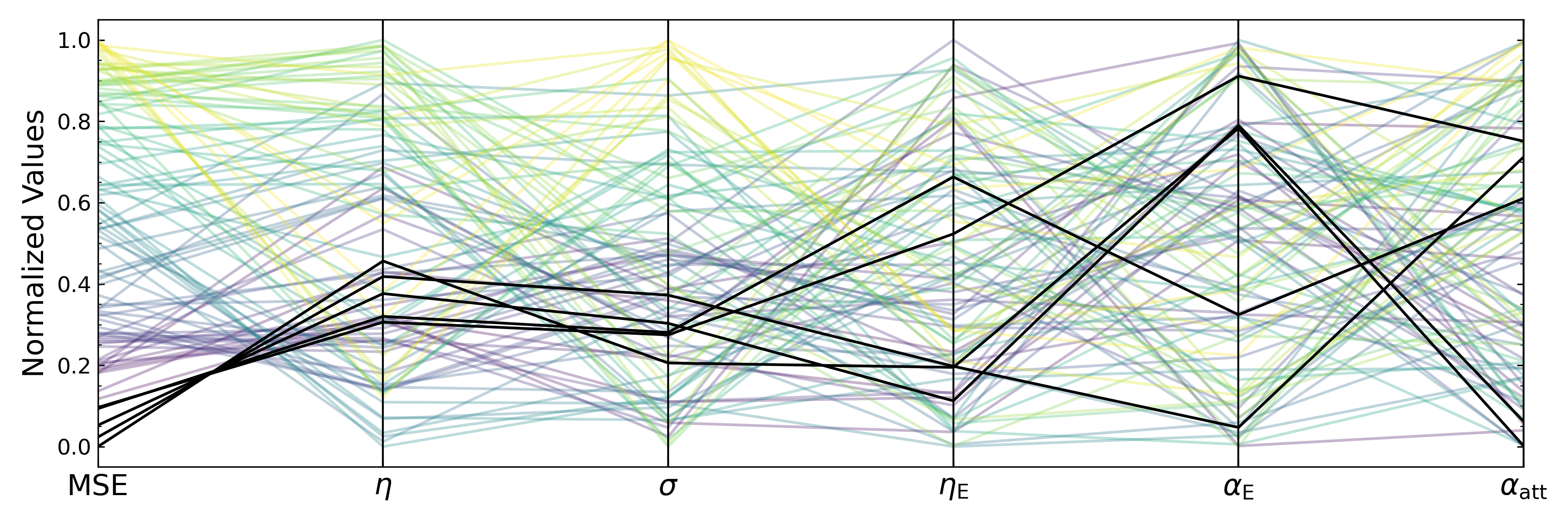}
    	\caption{Parallel coordinates plot summarizing hyperparameter optimization results for the small proxy model. Each line represents one of the 100 training runs, illustrating how different hyperparameter combinations affect model performance. The five best-performing runs are highlighted in black. The displayed hyperparameters are the learning rate ($\eta$; sampled from $10^{-4}$ to $10^{-1}$), the global initialization scale ($\sigma$), the embedding learning rate scale factor ($\eta_\text{E}$), and the embedding and attention scaling factors ($\alpha_{\text{E}}$ and $\alpha_{\text{att}}$), each sampled from the range [0.1, 10]. All axes are shown on a logarithmic scale and normalized between 0 and 1. The lines are colored according to their MSE values, with cooler colors (purples and blues) representing lower MSE values and warmer colors (yellows) representing higher MSE values.}
        \label{fig:fig2_hp_parallel_coords}
    \end{figure*}
\end{center}

Figure \ref{fig:fig2_hp_parallel_coords} shows a parallel coordinates plot summarizing the results of this hyperparameter search for the small proxy model. In this visualization, each colored line represents one training run, with its path across the six axes showing the specific combination of hyperparameter values and the resulting validation MSE. The five best-performing runs (those with lowest MSE) are highlighted in black. The colors of the lines appear to correspond to the MSE values, with cooler colors (purples and blues) representing lower MSE values and warmer colors (yellows) representing higher MSE values. This representation allows us to identify patterns and correlations between different hyperparameters and model performance.

The plot shows that the learning rate ($\eta$) and global initialization scale ($\sigma$) appear well-constrained among the best-performing models, with successful configurations clustering in relatively narrow ranges. The remaining three hyperparameters ($\eta_\text{E}$, $\alpha_{\text{E}}$, and $\alpha_{\text{att}}$) show greater variability among high-performing models, suggesting they are less critical for optimization. Interestingly, Fig.\,\ref{fig:fig2_hp_parallel_coords} suggests an anti-correlation between embedding and attention scaling factors, indicating that as one increases, the other typically decreases in high-performing models. This suggests a compensatory relationship where the relative balance between these factors matters more than their absolute values.

\begin{figure*}[!t]
    \centering
    \includegraphics[width=\linewidth]{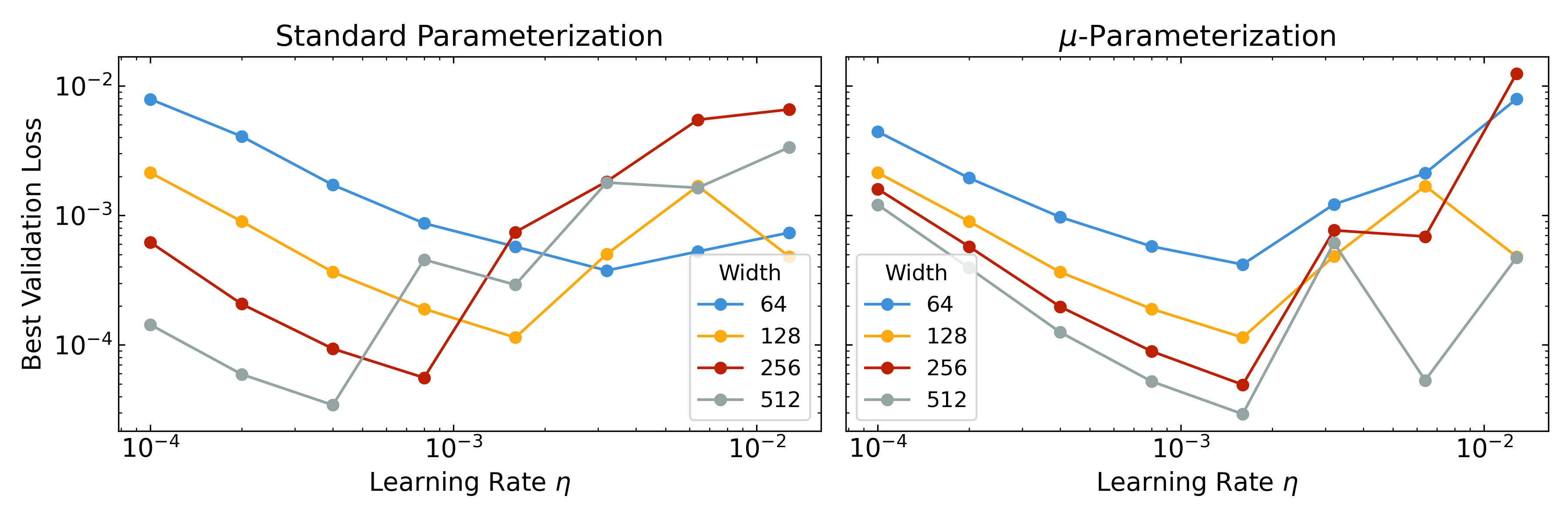}
    \caption{$\mu$P and Standard Parametrization exhibit different behaviors when scaling model width. Each plot shows validation loss (MSE) on the y-axis versus learning rate on the x-axis for models of varying widths (64, 128, 256, and 512 dimensions). Standard Parametrization (left panel) refers to conventional weight initialization without dimension-dependent scaling, while $\mu$P (right panel) applies specific scaling factors to weight initialization and learning rates based on matrix dimensions. The left panel shows that, under Standard Parametrization, the optimal learning rate (indicated by the minimum point of each curve) decreases as width increases, with the optimal learning rate shifting from approximately $3 \times 10^{-3}$ for width 64 to $4 \times 10^{-4}$ for width 512, a factor of approximately 3.2 decrease. The right panel highlights the stabilizing effect of $\mu$P on the optimal learning rate, with all model widths achieving their best performance at approximately $2 \times 10^{-3}$, demonstrating that our $\mu$P implementation successfully maintains consistent training dynamics across models of different scales. Additionally, note that wider models (higher dimensions) generally achieve lower validation loss values, reflecting increased model capacity.}
    \label{fig:fig3_mup_vs_sp_width}
\end{figure*}

\begin{figure*}[!t]
    \centering
    \includegraphics[width=\linewidth]{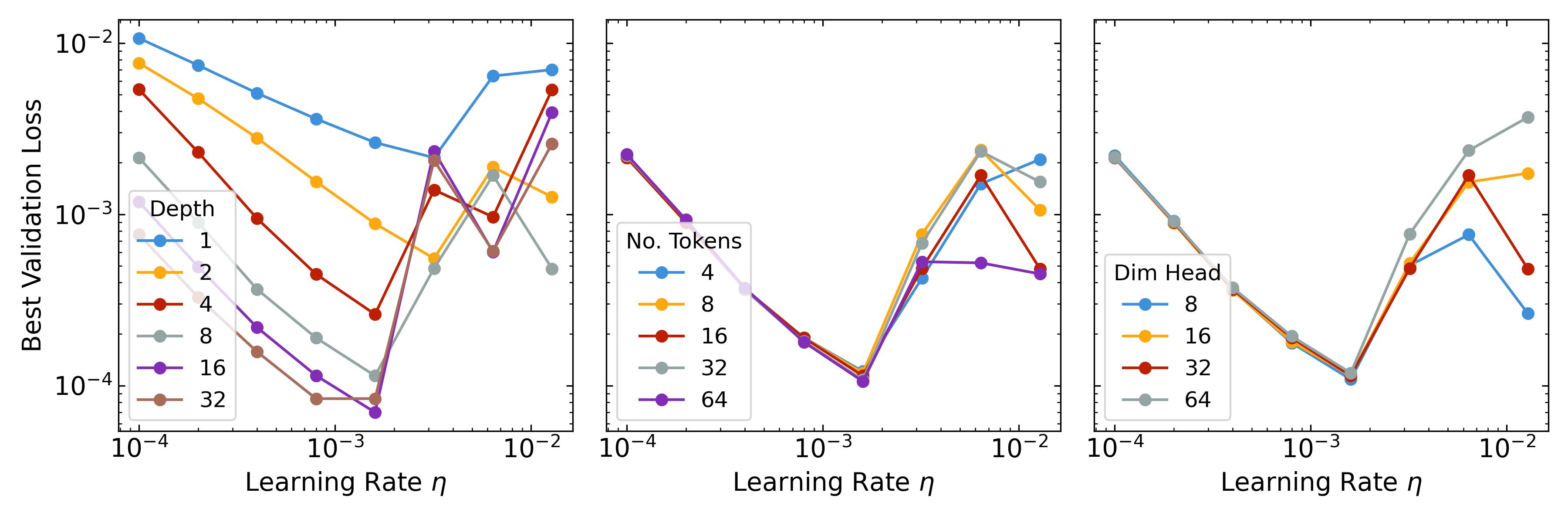}
    \caption{Stability of the optimal learning rate across different architectural dimensions: depth (number of Transformer blocks, left), number of parameter tokens (middle), and head dimensionality (number of attention heads, right). Each plot shows validation loss (MSE) versus learning rate for models with varying values of a single architectural parameter while keeping other parameters fixed. The optimal learning rate is indicated by the minimum point of each curve. Our $\mu$P implementation successfully maintains consistent optimal learning rates when varying the number of tokens and head dimensionality, with all curves reaching their minima at similar learning rate values (approximately $2\times10^-3$). For depth scaling (left panel), stability is maintained within approximately 0.5 to 2 times the proxy model's depth (8 layers), beyond which some retuning becomes beneficial as curves for very shallow (1-2 layers) and very deep models (16-32 layers) show shifted optima. This confirms that hyperparameters optimized on our proxy model transfer well across most architectural variations, with depth requiring more careful consideration.}
    \label{fig:fig4_mup_hyperparameters_stability}
\end{figure*}

This optimization process is a crucial first step in our scaling law investigation. By identifying optimal training hyperparameters on a small proxy model, we establish a baseline configuration that, according to $\mu$-Parametrization theory, should remain effective as we scale up model size in a principled manner. This approach avoids the prohibitive computational cost of repeating hyperparameter searches for each model scale, allowing us to focus on investigating how performance scales with model size, training data, and compute while maintaining optimal training dynamics.

After evaluating all runs, we identified the three best-performing sets of hyperparameters and computed their geometric mean to derive our final optimized values. The resulting hyperparameters were approximately as follows: learning rate $\eta = 0.0018$, global initialization scale $\sigma = 0.3890$, embedding learning rate scaling factor $\eta_\text{emb} = 0.2630$, embedding scaling factor $\alpha_{\text{E}} = 1.230$, and attention scaling factor $\alpha_{\text{att}} = 0.3236$. In the following experiments, these hyperparameters are fixed to the optimized values specified above, unless stated otherwise.

\subsection{\texorpdfstring{$\mu$}{mu}P Scaling Validation Tests} 
As the next step, we evaluated our implementation of $\mu$P by comparing it with the default Standard Parametrization and examining how the minimum validation loss varies across different learning rates. We trained a variety of models, each varying along a single dimension (i.e., width, depth, number of embedding tokens, or head dimensionality), while keeping all other hyperparameters fixed to match the reference proxy model described in the previous section ($d=128$, $N=8$, $t=16$, and $h=4$). 

This approach allows us to assess how effectively the optimal learning rate remains stable across these architectural parameters, which is particularly relevant when scaling model size. For all experiments, we used a cosine learning-rate schedule consisting of a 5,000-step warm-up followed by a cosine decay over 45,000 steps.

\begin{center}
    \begin{figure*}[!t]
        \centering
    	\includegraphics[width=0.8\hsize]{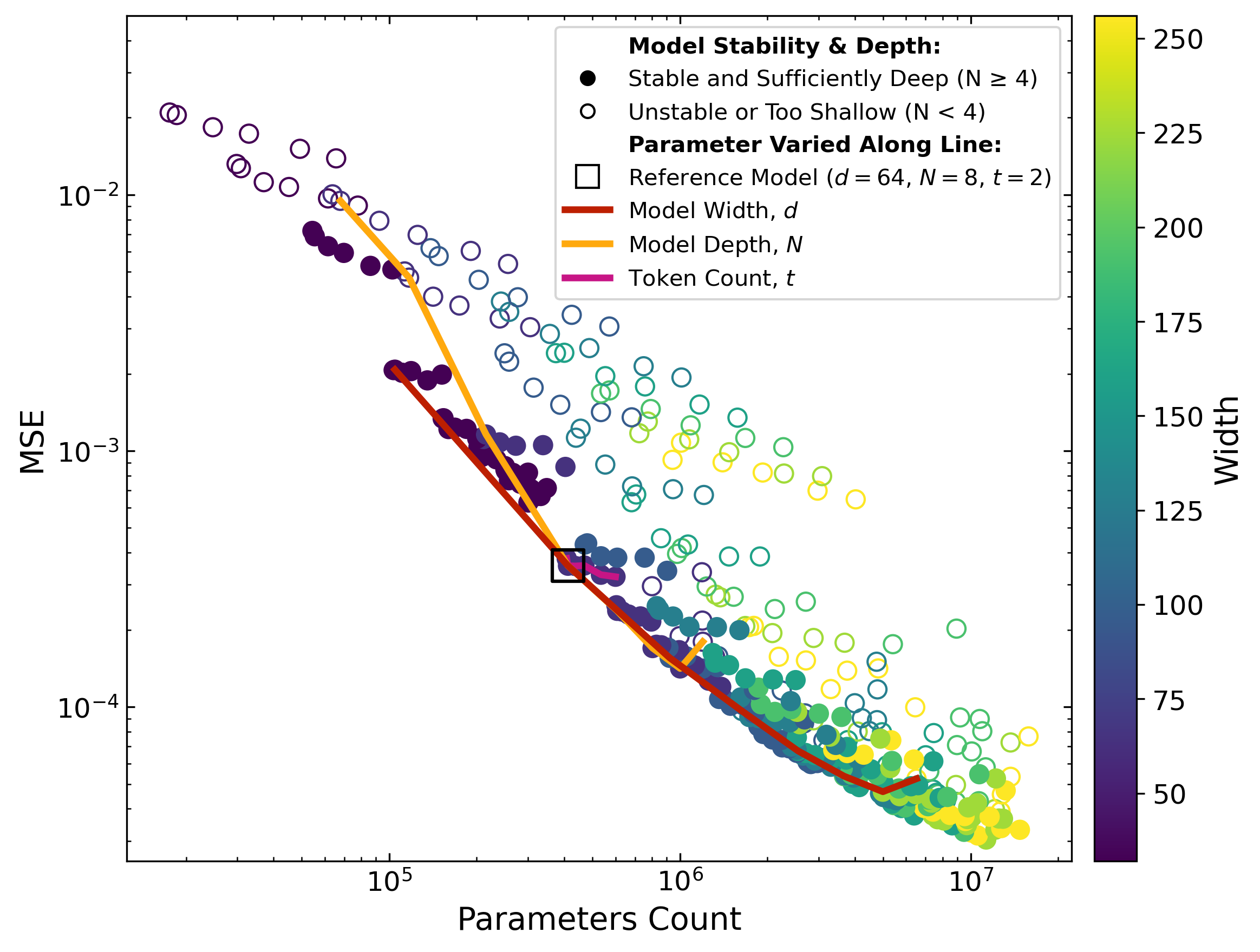}
    	\caption{Validation loss (MSE) of trained models across varying depths ($N$), widths (dimension $d$), and embedding token counts ($t$). Filled markers (167 models) indicate stable models with sufficient depth ($N \ge 4$) that converged during training, whereas open markers (195 models) indicate models either too shallow ($N < 4$), unstable, or both (96 shallow and 54 unstable). Stable models (filled markers) align along a narrow region, indicating that total parameter count is a strong predictor of model performance, while the specific depth-to-width ratio has less impact. Lines illustrate the effect of independently varying a single parameter from a reference model with shape $(d=64, N=8, t=2)$. Increasing model width consistently improves performance until signs of possible instability at the largest widths; shallow models perform significantly worse compared to deeper models with similar parameter counts; increasing depth beyond $N\ge4$ initially mirrors the benefit of increasing width but leads to instability sooner than scaling width; varying token count is notably less parameter-efficient than adjustments to depth or width. For a more detailed illustration of how each individual dimension influences emulation quality, see Fig.~\ref{fig:fig9_scaling_tpayne_shape_detail} in Appendix~\ref{appendix:scaling_tpayne}.
        }
        \label{fig:fig5_scaling_tpayne_shape}
    \end{figure*}
\end{center}

Figure~\ref{fig:fig3_mup_vs_sp_width} shows the results obtained from scaling the model width (embedding dimension $d$). The comparison reveals a crucial difference between the two parameterization approaches: under Standard Parametrization (left panel), which uses the conventional approach of initializing weights with a fixed variance regardless of layer size, the optimal learning rate shifts leftward (decreases) as model width increases, requiring recalibration for each model size. In contrast, with $\mu$P (right panel), which scales initialization and learning rates according to layer dimensions, the optimal learning rate remains relatively stable across different widths, with the loss curves for various model sizes reaching their minima at approximately $2 \times 10^{-3}$, regardless of whether the model width is 64 or 512.

In Fig.\,\ref{fig:fig4_mup_hyperparameters_stability}, we present results for several other architectural hyperparameters (depth, number of tokens, and head dimensionality) and examine their impact on the optimal learning rate under $\mu$P. The results demonstrate that $\mu$P successfully stabilizes the optimal learning rate across multiple architectural dimensions. Varying the number of tokens (middle panel) and head dimensionality (right panel) had negligible influence on the optimal learning rate, indicating that hyperparameters optimized on our proxy model transfer excellently across models with different token counts and attention head configurations.

For depth scaling (left panel), $\mu$P maintains stability within a range of approximately 0.5 to 2 times that of the proxy model used for hyperparameter optimization. While this demonstrates good transferability within a moderate depth range, it suggests that for more extreme depth variations, some retuning of hyperparameters may be beneficial. This observation aligns with previous research indicating that depth scaling presents unique challenges compared to width scaling in Transformer architectures.

\subsection{Architectural Parameter Optimization for Transformer Scaling}
\label{sec:results_small_models}

Having established the effectiveness of our $\mu$P implementation for stabilizing learning rates across model widths, we next sought to determine the optimal allocation of parameters across different architectural dimensions. This investigation is crucial for developing efficient scaling strategies, as it reveals which components contribute most effectively to model performance as total parameter count increases.

We conducted an extensive grid search to determine how best to scale width ($d$), the number of layers ($N$), and the number of tokens ($t$) for a given number of parameters in a TransformerPayne model. When increasing the width, we kept the head dimension fixed at 32, resulting in the number of heads $h = d / 32$. As shown in the left panel of Fig.\,\ref{fig:fig4_mup_hyperparameters_stability}, head dimensionality did not significantly impact the minimum training loss within the explored range. Since preliminary experiments indicated that models with larger depths frequently exhibit signs of instability, we introduced an additional instability criterion: a model was flagged as unstable if its validation loss, measured by mean absolute error (MAE) and evaluated every 2,500 training steps, increased by more than a factor of two between any two consecutive evaluations.

The architectural parameters were varied systematically on a regular grid: width ($d$) was set to [32, 64, 96, 128, 160, 192, 224, 256], the number of layers ($N$) to [1, 2, 4, 8, 12, 16, 20, 24], and the number of tokens ($t$) to [1, 2, 8, 16, 32, 48]. This resulted in 384 training runs, of which 362 were completed successfully. Each training run used a training dataset consisting of 100,000 spectra and lasted 50,000 steps, following a WSD schedule consisting of 5,000 warm-up steps and a 10,000-step cool-down phase. Since WSD schedules typically require a lower optimal learning rate compared to a linear warm-up cosine decay schedule, we reduced the learning rate from 0.0018 to 0.0012 \citep[see][]{2025arXiv250118965S}.

To further stabilize training, we applied gradient clipping with a global norm threshold of 1.0. Specifically, if the global gradient norm exceeded 1, it was rescaled to exactly 1 without altering its direction.

The results of these scaling experiments are summarized in Fig.\,\ref{fig:fig5_scaling_tpayne_shape} (see also Fig.\,\ref{fig:fig9_scaling_tpayne_shape_detail} in Appendix \ref{appendix:scaling_tpayne}). Models indicated by filled markers (167 models) exhibited convergence during training and are sufficiently deep ($N \ge 4$). Models with depth $N < 4$ systematically showed poorer performance and distinct training dynamics compared to deeper models ($N \ge 4$). We thus adopted this threshold to separate shallow from deeper models. This distinction held consistently across the explored widths, indicating a general change in model behavior as depth increases. In contrast, open markers (150 models) represent configurations that were either too shallow ($N < 4$) or exhibited clear signs of instability, as defined above. Among these open markers, 96 correspond to shallow models and 54 to unstable models.

The stable models align along a narrow region, indicating that the total parameter count strongly predicts model performance, whereas the specific depth-to-width ratio is less critical within the wide range of explored proportions. The lines illustrate the effect of independently varying individual parameters from a reference configuration ($d = 64$, $N = 8$, $t = 2$). Increasing the model width consistently enhances performance, although the gains diminish at larger widths. Shallow models exhibit worse performance compared to deeper models with similar parameter counts. 

Increasing depth beyond $N \ge 4$ initially yields performance improvements comparable to scaling width, but leads to instability at smaller total parameter counts than width scaling does. This behavior is expected, as illustrated in the leftmost panel of Fig.\,\ref{fig:fig4_mup_hyperparameters_stability}, which shows that $\mu$P does not fully stabilize the optimal learning rate as model depth increases. Finally, increasing the token count is substantially less parameter-efficient than adjusting either depth or width. Based on these findings, in our subsequent scaling experiments, we will prioritize increasing model width while ensuring sufficient depth for effective representation learning, as this approach offers the most stable and parameter-efficient path to improved performance.

\subsection{Scaling TransformerPayne}
\label{sec:results_scaling}

Having optimized our training hyperparameters and established guidelines for scaling model architecture, we are now positioned to systematically investigate how model performance scales with increasing resources. We trained ten TransformerPayne models to comprehensively explore scaling behavior across multiple dimensions. We fixed the token count ($t=16$) and head dimensionality ($h=32$) across all models as our previous experiments showed these parameters had minimal impact on performance compared to model width and depth. Instead, we systematically varied width ($d$) and depth ($N$) as these dimensions offered the most efficient paths to improved performance. The models were designed so that each successive model approximately doubled in parameter count relative to the previous one. The resulting architectures and parameter counts are summarized in Table~\ref{tab:model_sizes}.

\begingroup
\setlength{\tabcolsep}{12pt}
\begin{table}[ht!]
\centering
\caption{Summary of TransformerPayne models trained}
\begin{tabular}{c c c}
\hline
\textbf{Width ($d$)} & \textbf{Depth ($N$)} & \textbf{Parameters ($P$)} \\
\hline \hline
32   & 4   & $6.98\times10^{4}$ \\
32   & 8   & $1.19\times10^{5}$ \\
64   & 4   & $2.73\times10^{5}$ \\
64   & 8   & $4.69\times10^{5}$ \\
96   & 8   & $1.05\times10^{6}$ \\
128  & 8   & $1.86\times10^{6}$ \\
160  & 12  & $4.14\times10^{6}$ \\
224  & 12  & $8.10\times10^{6}$ \\
256  & 16  & $1.37\times10^{7}$ \\
384  & 16  & $3.09\times10^{7}$ \\
\hline
\end{tabular}
\parbox{0.5\textwidth}{\vspace{1em} \small Token count ($t=16$) and head dimensionality ($h=32$) were fixed for all models, while width ($d$) and depth ($N$) were systematically varied. Each successive model approximately doubles the parameter count of the previous one.}
\label{tab:model_sizes}
\end{table}
\endgroup

\begin{center}
    \begin{figure*}[!t]
        \centering
    	\includegraphics[width=1.0\hsize]{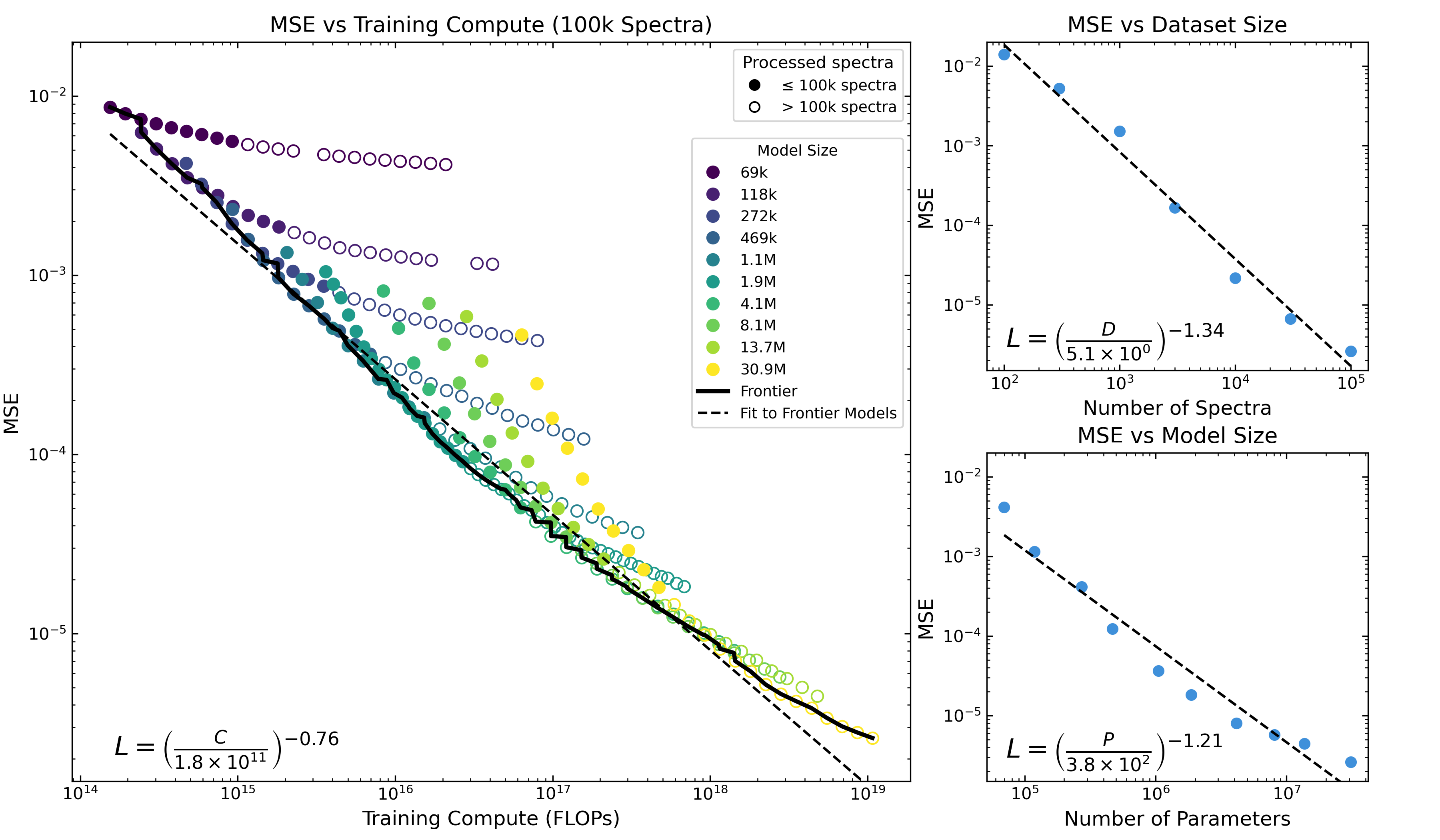}
    	\caption{Scaling behavior of TransformerPayne models. Dashed lines represent power-law fits with corresponding equations indicated in each panel. Left panel: MSE versus training compute (in FLOPs) for all model configurations using the full 100,000 spectra dataset. Each point represents a converged training run, with colors indicating model size (ranging from 69k to 30.9M parameters). Filled circles represent training runs that process fewer than 100,000 total examples (less than one epoch), while open circles indicate runs exceeding one epoch where examples are revisited. The solid black line traces the compute frontier, the optimal performance achievable at each compute level, while the dashed line shows the power-law fit to this frontier with exponent $-0.76$. Upper-right panel: MSE versus dataset size showing the best performance achieved for each dataset size across all model architectures and training durations. The $-1.34$ exponent indicates a 22-fold reduction in MSE for each 10-fold increase in training data. Lower-right panel: MSE versus model size showing the best results achieved across all dataset sizes and training durations for each model size. The power-law fit with exponent $-1.21$ demonstrates a 16-fold reduction in MSE per 10-fold increase in parameter count. Note the slight flattening for the largest models, suggesting they approach the limits imposed by the dataset size. Together, these panels illustrate distinct but complementary scaling relationships across the three primary dimensions: compute, data, and model capacity.
}
        \label{fig:fig6_scaling_laws_main}
    \end{figure*}
\end{center}

To investigate scaling behavior, we explored not only model size but also training dataset size and training duration. For the training set dimension, we created seven subsets of the full dataset, varying in size from 100 to 100,000 examples, thus exploring scaling across three orders of magnitude. To scale training compute, we varied the number of training steps according to:
\begin{equation}
S_n = 10^5 \cdot \left(\frac{5}{4}\right)^n,\quad\text{for } n \in \{-9, -8, \dots, 14\},
\end{equation}
rounded to the nearest ten. Consequently, the shortest runs lasted 13,420 steps, while the longest extended to 2,273,740 steps.

For all training runs, we used a learning rate of 0.0012 and a WSD schedule comprising 10,000 warm-up steps, followed by a stable phase of constant learning rate, and concluding with decay during the final 20\% of training steps. We fixed batch sizes across all experiments to facilitate consistent comparisons. To optimize computational efficiency, we adopted a checkpoint reuse strategy: a single, very long training run was executed with frequent intermediate checkpoints, from which we subsequently launched multiple shorter decay runs in parallel. This strategy significantly reduced the computational cost associated with exploring scaling laws by reusing common initial portions of the training runs. This approach enabled us to efficiently trace the entire compute frontier for a fixed dataset size and model, while varying only the training length. An ensemble of such training runs, corresponding to a selected fixed dataset size and model, is illustrated in Fig.\,\ref{fig:fig10_wsd_explained} in Appendix\,\ref{appendix:scaling_tpayne}.

\subsubsection{Scaling Dimensions and Optimization Strategy}

Experiments were focused on investigating three TransformerPayne scaling regimes as described in Sec.\,\ref{sec:methods_neural_scaling_laws}. These are about quantifying how emulation accuracy improves: when dataset size is fixed (with optimized choices for model size and compute allocation for that dataset size), when model size is fixed (with optimized choices for dataset size and compute allocation for that model) and when training compute is fixed (with optimized choices for dataset and model sizes). These relationships are captured by power-law relations as given in Equations \ref{eq:scaling_data_size}, \ref{eq:scaling_model_size}, and \ref{eq:scaling_compute}. By \textit{optimized choices}, we mean that, for each fixed resource (e.g., dataset size), we report the best performance achieved across all variations of the other explored resources, effectively showing the performance envelope achievable under our experimental constraints.

It is important to note that the optimal configuration in each scaling dimension does not necessarily correspond to maximum values in the remaining dimensions. For example, when fixing dataset size and studying the relationship between validation loss for different model sizes and compute resources, the largest model size does not always yield the best performance. This is because, at a fixed compute budget, larger models can only be trained for fewer steps, potentially leading to undertraining. The optimized choice therefore represents the best balance between the two unfixed dimensions while keeping the third dimension constant. The three dimensions we explored are model size (summarized in Table~\ref{tab:model_sizes}), dataset size (ranging from 100 to 100,000 examples), and training duration (following the step schedule defined in Sec.\,\ref{sec:results_scaling}). 

Thus, for all considered TransformerPayne sizes, spectral grid sizes, and training durations, we investigated how emulation quality could be maximized when varying one parameter at a time, while optimally selecting values for the other two parameters.

\subsubsection{Scaling with Dataset Size and Model Size}

The top-right panel of Fig.\,\ref{fig:fig6_scaling_laws_main} demonstrates scaling with dataset size, showing the best achieved MSE for each dataset size across all model architectures and training durations. This analysis reveals a power-law relationship with exponent $-1.34$, indicating a 22-fold reduction in MSE whenever the dataset size increases by a factor of 10. This quantifies the substantial benefit provided by additional training data. Notably, as illustrated by the left panel, performance with the largest dataset (100,000 spectra) continues to improve with increased compute, suggesting even better results could be achieved by extending training further. For smaller datasets, model performance has already saturated, and the largest models begin to overfit; additional examples of this behavior are presented in Fig.\,\ref{fig:fig11_raw_scaling_examples} in Appendix \ref{appendix:scaling_tpayne}.

The bottom-right panel of Fig.\,\ref{fig:fig6_scaling_laws_main} demonstrates scaling with model size, displaying the best results across all dataset sizes and training durations. The fitted power-law has an exponent of $-1.21$, indicating a 16-fold reduction in MSE per 10-fold increase in parameter count. A slight flattening is observed for models larger than approximately $10^7$ parameters, suggesting that these models are approaching the performance limit imposed by the current dataset size (100,000 spectra). Excluding the two largest models yields a steeper exponent of $-1.41$, and excluding the three largest yields an exponent of $-1.54$.

\subsubsection{Compute Frontier Analysis}

The left panel of Fig.\,\ref{fig:fig6_scaling_laws_main} shows how MSE varies as a function of the compute budget allocated to training. Unlike the right panels which show optimized performance across dimensions, this plot displays the raw performance trajectory of all model configurations across varying amounts of compute. Each point represents a converged training run, with model size indicated by color (purple for the smallest models, transitioning to yellow for the largest).

The distinction between filled and open circles represents a fundamental transition in training dynamics: filled circles indicate training runs where the total number of processed examples (batch size $\times$ training steps) is less than 100,000, meaning the model has not completed a full pass through the dataset (less than one epoch). Open circles represent runs exceeding one epoch, where the model has seen some training examples multiple times. This visualization shows how performance continues to improve even after multiple passes through the dataset, though with diminishing returns compared to the sub-epoch regime.

The solid black line traces the compute frontier --- the lower envelope of all points representing the best achievable performance at each compute level regardless of model size. This frontier represents the optimal trade-off between model size and training duration for each compute budget. Notably, for compute budgets exceeding approximately $10^{17}$ FLOPs, the frontier transitions into the multi-epoch regime (open circles), indicating a potential shift in scaling efficiency. Extending beyond one epoch may decrease efficiency since additional compute is used to revisit already-seen examples rather than exploring new regions of the parameter space through increased model capacity or novel training examples. This reuse of training data introduces diminishing returns as the model has already extracted much of the unique information available in these examples. 

A power-law fit to all models along this frontier yields an exponent of $-0.76$, indicating that a 10-fold increase in compute reduces the MSE approximately by a factor of 5.7. In contrast, for compute budgets below $10^{17}$ FLOPs (where training on the frontier remains within one epoch), scaling is more favorable, with an exponent of $-0.87$. This difference highlights that in the sub-epoch regime, each additional computation contributes to learning from new, unseen data, creating a more efficient scaling relationship. Thus, only in this lower-compute regime does the frontier strictly trace what would be expected in an infinite-data limit where example repetition is unnecessary. 

\begin{center}
    \begin{figure*}[!t]
        \centering
    	\includegraphics[width=1.0\hsize]{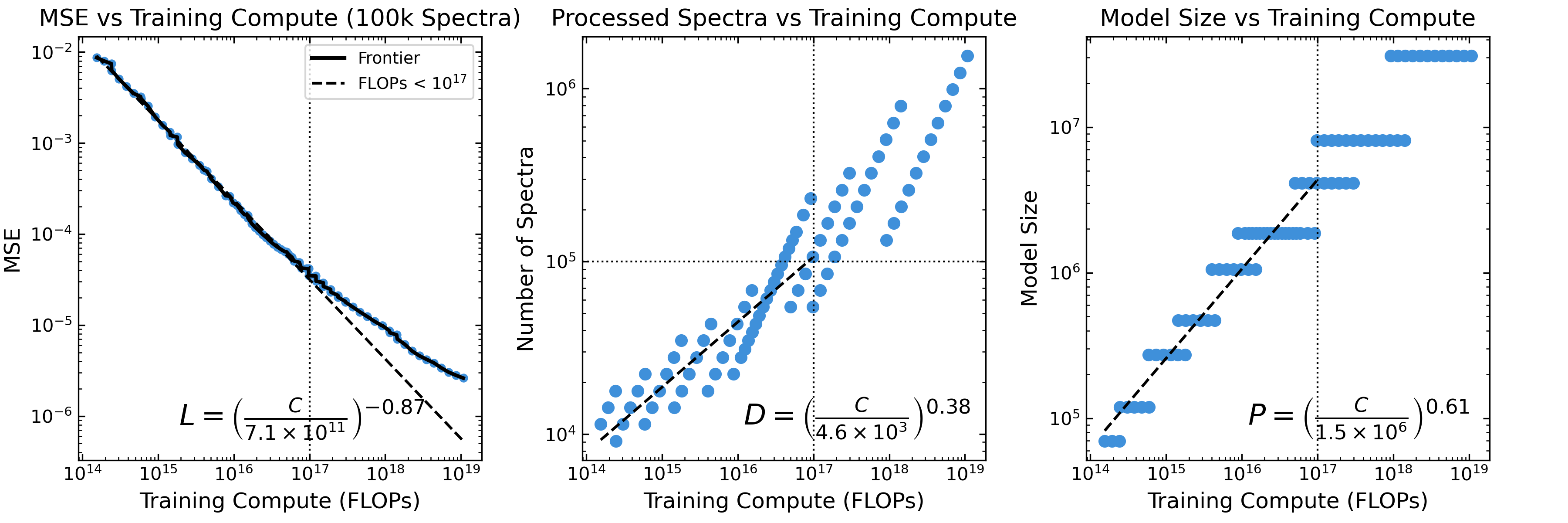}
    	\caption{TransformerPayne models scaling along compute frontier. Dashed lines represent power-law fits with corresponding equations indicated in each panel. Left panel: Validation loss (MSE) versus compute, focusing specifically on the frontier of best-performing models (the lowest achievable MSE at each compute level). Unlike Fig.\,\ref{fig:fig6_scaling_laws_main}, this panel isolates only the optimal models and highlights the power-law fit for the sub-epoch regime (training compute $\leq 10^{17}$ FLOPs), where scaling behavior is most reliable. Middle panel: Number of processed training examples versus compute for models along the frontier. Each blue dot represents a model on the compute frontier, with diagonal patterns formed by individual model architectures trained for increasing durations. The power-law fit with exponent 0.38 indicates that when compute increases 10-fold, the optimal training set size should increase approximately 2.4-fold. The horizontal dotted line at 100,000 spectra marks our dataset size limit; points above this line represent training beyond one epoch. Right panel: Model size (number of parameters) versus compute for frontier models. The horizontal bands correspond to discrete model architectures from Table~\ref{tab:model_sizes}, with transitions between bands indicating compute thresholds where larger models become more efficient than smaller ones trained longer. In all panels, vertical dotted lines at $10^{17}$ FLOPs indicate the boundary beyond which our dataset becomes the limiting factor, and all power-law fits are restricted to the region left of this boundary to ensure reliable scaling estimates.
}
        \label{fig:fig7_scaling_laws_supp}
    \end{figure*}
\end{center}

\subsubsection{Optimal Resource Allocation}

While the right panels of Fig.\,\ref{fig:fig6_scaling_laws_main} show how MSE scales with dataset size and model size independently, what is often more practically valuable is understanding how to optimally balance resources when compute is the primary constraint. This point is particularly important because the largest models are not always optimal. Under limited compute budgets, smaller models trained for more optimization steps often outperform larger models trained for fewer steps. Fig.\,\ref{fig:fig7_scaling_laws_supp} addresses precisely this practical question.

Focusing more closely on the model and dataset sizes associated with the compute frontier, we interpret the results presented in Fig.\,\ref{fig:fig7_scaling_laws_supp}. The left panel revisits the relationship between MSE and training compute, but unlike the previous figure, here we highlight only the frontier models, those achieving the best performance at each compute level through optimized choices of model size and training duration. For each compute budget, this represents the single best-performing configuration from among all the models we trained.

The left panel restricts our power-law fit to the region below $10^{17}$ FLOPs (indicated by the vertical dotted line), where training consistently remains within one epoch. In this regime, each training example is seen at most once, and our data is not constrained by the finite size of our stellar spectra grid. This yields a defined power-law exponent of $-0.87$. Within this regime, a 10-fold increase in compute reduces MSE by a factor of approximately 7.4, demonstrating that with properly optimized architectural choices, compute scaling can be substantially more efficient than suggested by the overall trend. 

Importantly, achieving this optimal scaling requires selecting the right model size for each compute budget, too large a model with insufficient training steps or too small a model with excessive training both result in suboptimal performance. Similarly, the number of training examples must be carefully balanced; as we show in the middle panel, this optimal number increases with available compute but at a rate slower than the growth in model size.

The middle panel shows how the optimal number of training examples scales with compute. Each blue dot represents a model on the compute frontier, plotting the total processed spectra against the compute used. The distinct diagonal patterns visible in this plot emerge because each diagonal line represents a single model architecture trained for increasing numbers of steps. As we move along any individual diagonal, we're seeing the same model trained for progressively more steps, processing more total examples and using more compute. The frontier (best-performing models) jumps between these diagonal lines, transitioning to larger models as compute increases, creating the stair-step pattern. If we had the computational resources to train models with continuously varying numbers of parameters (rather than our discrete set), we would expect the optimal frontier to follow the fitted dashed line more smoothly. 

We strategically limited our power-law fit to the regime below $10^{17}$ FLOPs where the processed spectra remain within one epoch (below the 100,000 spectra dotted line), as this represents the most reliable scaling regime where each new computation contributes to processing previously unseen data. Since our results indicate it is optimal to avoid processing spectra more than once, this relationship also directly suggests the ideal training dataset size needed for a given compute budget, approximately the number of processed spectra indicated by the dashed line. The resulting power-law has an exponent of 0.38, meaning that as compute increases by a factor of 10, the optimal number of processed examples should increase roughly by a factor of 2.4. The horizontal dotted line at 100,000 spectra indicates our current dataset limit; points above this line represent models trained over multiple epochs, where the same examples are revisited.

The right panel reveals how model size should scale with increasing compute budgets. Each blue dot represents a model on the frontier, showing which model size performs best at each compute level. Similar to the middle panel, we observe distinct horizontal bands corresponding to our discrete set of model architectures (from Table \ref{tab:model_sizes}). As compute increases, there are clear transition points where it becomes more efficient to switch from a smaller model trained for many steps to a larger model trained for fewer steps. 

The progression of these transitions reveals a consistent pattern that follows a power-law with an exponent of 0.61, implying that when compute increases by a factor of 10, optimal model size should grow by approximately a factor of 4.1. If we had tested a continuous spectrum of model sizes, we would expect the actual frontier to follow the dashed line more smoothly. Again, we strategically fit only points below $10^{17}$ FLOPs to remain in the single-epoch regime where scaling behavior is most reliable.

These findings deliver distinct but complementary insights compared to Fig.\,\ref{fig:fig6_scaling_laws_main}. While the earlier figure characterized the theoretical scaling potential of each dimension in isolation (showing how performance improves when scaling just dataset size or just model size with exponents $-1.34$ and $-1.21$ respectively), Fig.\,\ref{fig:fig7_scaling_laws_supp} provides practical guidance for real-world resource allocation under compute constraints.

\section{Discussion}
\label{sec:discussion}

In this study, we investigated how the emulation accuracy of TransformerPayne scales with increases in stellar spectra grid size, training compute budget, and neural network size. Our experiments demonstrate that, given stable training and careful scaling of hyperparameters, emulation accuracy predictably improves across several orders of magnitude following consistent power-law relationships. Specifically, we find that optimal scaling involves simultaneously increasing training compute, dataset size, and model size according to well-defined proportions. These scaling laws, which closely parallel those established in language modeling provide practical guidance for optimal resource allocation in spectral emulation. They offer a quantitative framework for designing stellar grids and planning computational requirements when targeting specific precision goals for current and future stellar surveys.

\subsection{Toward Spectral Foundational Models}

Developing spectral foundational models for stellar spectrum emulation presents two main challenges. First, stable and efficient training must be maintained across varying model sizes and training durations. Second, determining the optimal way to scale network capacity is crucial, as multiple hyperparameters (e.g., width, depth, and number of tokens in TransformerPayne) simultaneously influence the total parameter count.

As demonstrated in this study, $\mu$-Parametrization proved to be a highly valuable tool for simplifying the scaling process, ensuring robust training dynamics when adjusting model width. While $\mu$P has been primarily demonstrated for language models using cross-entropy loss, our experiments confirm its effectiveness in the regression context of stellar spectra emulation with MSE loss. $\mu$P reliably stabilized scaling with respect to width; however, we found that optimal hyperparameters required adjustments when varying depth by more than a factor of two compared to the reference model.

Embedding token count and head dimensionality had only minor effects on optimization stability. These findings align closely with broader empirical results \citep[e.g.,][]{2024arXiv240405728L}, suggesting that hyperparameters remain stable primarily with width but require more careful retuning for significant changes in depth. Therefore, future hyperparameter searches would benefit from explicitly tuning at multiple reference depths (e.g., $N=2$, $N=8$, and $N=32$) to improve stability when scaling depth substantially. Despite this limitation, our results indicate that with sufficient depth, our current hyperparameter settings remain nearly optimal across a wide range of architectures.

Finally, we found that, for many width-depth combinations, models of similar total parameter counts achieve comparable performance, provided they are not too shallow. Specifically, models with fewer than four Transformer blocks reliably underperformed compared to deeper architectures of similar size. In practice, once a model is sufficiently deep and training is stable, performance primarily depends on the total parameter count rather than the exact depth-to-width ratio.

Thus, maintaining training stability ultimately proved more critical than fine-tuning exact depth-to-width ratios, especially since instabilities became more frequent at greater depths without careful re-optimization of training-related hyperparameters. This finding simplifies the architectural design process, as it suggests that, beyond ensuring sufficient depth, practitioners can focus more on total parameter count through increasing width than on precise architectural proportions when scaling TransformerPayne models.

\subsection{Practical Implications for Spectral Emulation}

Neural-network emulators are typically trained under various computational and data constraints, with the overarching goal of minimizing the impact of emulation errors on subsequent inference of stellar parameters and abundances. In this study, we adopted plane-parallel LTE atmospheric models, but we expect the scaling relationships we have described to remain qualitatively robust across different modeling approaches, even as the specific quantitative constants may vary. Our methodology should serve as a template for future spectroscopic emulation efforts.

When developing spectral emulators, the central challenge is optimizing performance within finite computational resources. Our analysis reveals that focusing exclusively on any single dimension, larger models, more training data, or extended training, yields suboptimal results. Instead, a balanced approach is necessary, where these resources scale together according to specific proportions.

The key practical scenario of interest is when a specific emulation precision (associated with target MSE) is required. In such cases, one should determine the necessary compute budget and then appropriately scale the model size and training dataset size according to the relationships illustrated in Fig.\,\ref{fig:fig7_scaling_laws_supp}. Specifically, given a desired validation loss, the required compute $C$ can be estimated by inverting the fitted scaling relation:
\begin{equation}
    C = C_c\,\mathcal{L}^{-1/\alpha_C},
\end{equation}
with parameters $C_c = 7.1\times10^{11}$ and $\alpha_C=0.87$. Subsequently, the dataset size $D$ and model size $P$ should follow from the computed budget as:
\begin{equation}
    D = \left(\frac{C}{4.6\times10^3}\right)^{0.38},\quad
    P = \left(\frac{C}{1.5\times10^6}\right)^{0.61}.
\end{equation}

These relationships reveal the key insight of our work: as computational resources increase, both model size and dataset size should increase together, but at different rates. Specifically, when compute increases by a factor of 10, model size should increase by approximately 4$\times$ while dataset size should increase by about 2.4$\times$. This balanced scaling ensures optimal performance for a given computational budget.

To illustrate the practical value of these scaling laws, consider planning for a large-scale emulator with 1 billion parameters (a 1B TransformerPayne model). Using our scaling equations, we can determine that the required training compute would equal $8.52 \times 10^{20}$ FLOPs, and the spectral grid should contain approximately $3.64 \times 10^{6}$ stellar spectra. Following these guidelines, the resulting MSE for this model would be approximately $1.26 \times 10^{-8}$.

To contextualize the computational requirements, we can convert abstract FLOP counts into hardware-specific metrics. For instance, using a modern Nvidia H100 accelerator, which delivers approximately 67 teraFLOPS in Float32 precision, the training time would be:
\begin{equation}
\frac{8.52 \times 10^{20}}{67 \times 10^{12} \times 24 \times 3600} \approx 147.10~~\text{H100-days}.
\end{equation}

These concrete calculations demonstrate how our scaling laws can be applied to practical planning decisions, allowing researchers to estimate resource requirements for specific emulation precision targets. Similar estimations can be conducted for various model sizes, spectral grid sizes, or target MSE values, providing a quantitative framework for resource allocation in spectroscopic studies.

\subsection{Limits of Neural Scaling Laws}

\begin{center}
    \begin{figure}[!t]
        \centering
    	\includegraphics[width=1.0\hsize]{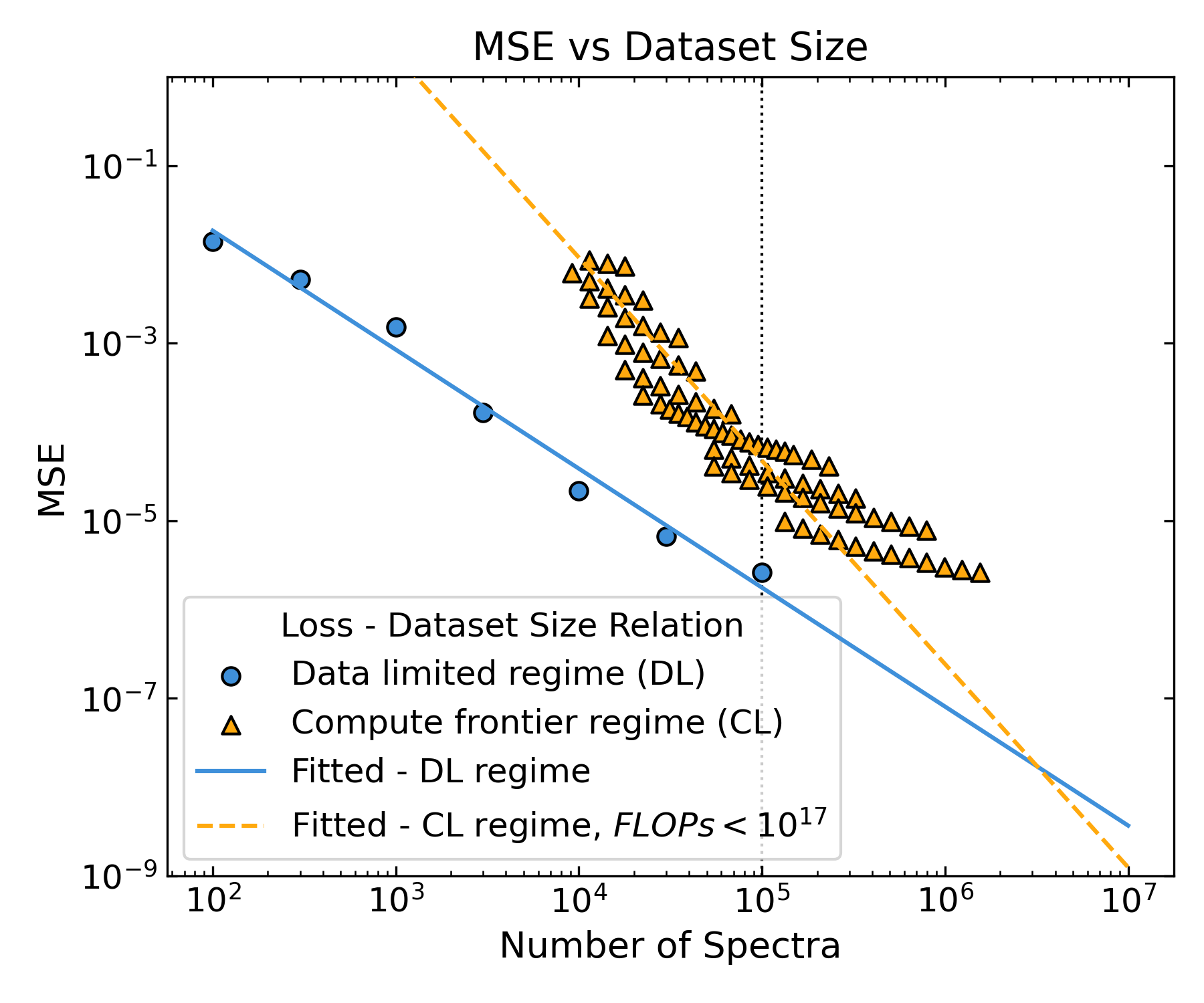}
    	\caption{Comparison of two different scaling regimes. The blue circles and line represent the \textit{data-limited regime}, showing how MSE decreases with more training data when using optimally-sized models trained to full convergence, regardless of computational efficiency. These points are derived from selecting the best results for each dataset size across all model architectures and training durations in our experiments. In contrast, the orange triangles represent the \textit{compute frontier regime}, showing how MSE scales with the number of processed spectra when using models that are optimal under constrained computational resources. Each triangle indicates a distinct converged model characterized by specific combinations of model size and training duration, with associated compute budget measured in FLOPs. The vertical dotted line indicates the size of our current spectral grid (100,000 spectra). The intersection of these trends at approximately $3\times10^6$ spectra indicates a transition point where our compute-optimal scaling approach reaches the data-limited ceiling. 
        }
        \label{fig:fig8_dataset_conjecture}
    \end{figure}
\end{center}

While empirical scaling laws suggest that model performance continues to improve with larger models, more training data, and increased training compute, extending these laws too far reveals apparent contradictions, indicating that the outlined scaling relationships must eventually break down. This point becomes particularly evident when comparing the two distinct scaling regimes illustrated in Fig.\,\ref{fig:fig8_dataset_conjecture}.

The first regime, represented by the blue line and circles, illustrates the theoretical best-case performance achievable with increasing amounts of training data. This \textit{data-limited regime} corresponds exactly to the scenario previously presented in the lower-right panel of Fig.\,\ref{fig:fig6_scaling_laws_main}, where models are trained to saturation to fully extract all available information from the training data. 

Unlike the compute frontier models, which aim to find the optimal balance between model size and training duration under compute constraints, these data-limited points represent exhaustive training to convergence, regardless of computational efficiency. The points along this line are directly reproduced from our earlier experiments, where we selected the best results achieved for each dataset size across all model architectures and training durations.

The second regime, represented by the orange triangles and dashed line, illustrates how performance scales along the compute frontier. In this \textit{compute frontier regime}, each orange triangle represents a distinct model that achieved optimal performance given its particular compute budget. These data points are identical to those presented previously in Fig.\,\ref{fig:fig7_scaling_laws_supp}, but here they are shown as MSE versus the number of training spectra rather than versus training compute.

These two trends intersect at approximately three million spectra, marking a potential transition in scaling behavior. The blue line represents the theoretical limit if one could fully exploit a given grid of stellar spectra without restrictions imposed by model size or computational resources. Validation loss cannot be improved beyond this theoretical limit (blue line) for a given dataset size. At the intersection point, our compute-optimal scaling approach (orange triangles) reaches the data-limited ceiling (blue line), indicating that the scaling laws we have established in this study would no longer apply beyond this point.

This does not mean that performance improvement stops entirely, rather, it would continue to improve but at the slower rate dictated by the data-limited regime (blue line) instead of following our compute-optimal scaling trajectory. Based on our scaling equations, this transition occurs at approximately 3 million spectra, an MSE of roughly $10^{-8}$, and corresponds to a model size of around $10^{9}$ parameters (1B parameters) as determined by our compute-optimal scaling equation.

Similar observations have been reported in large-scale language modeling studies by \citet{2020arXiv200108361K}, who interpreted this phenomenon as a transition between two distinct regimes: one where increasing data reliably enhances performance, and another where the available data is effectively exhausted. The authors suggest that scaling laws must inevitably break around these transition points, hypothesizing a deeper fundamental implication: that at these points, neural networks have fully extracted the information available in a training corpus, and the corresponding loss could serve as an estimate of the entropy-per-token in natural language.

Although a dataset with stellar spectra is not inherently probabilistic, some randomness might still be introduced by two factors: first, by the numerical modeling code used to simulate stellar atmospheres (in this case, Kurucz's Atlas and Synthe) and its limited numerical precision; second, by the linear interpolation process in wavelength that we use to sample flux at arbitrary wavelengths. The numerical precision of Kurucz models typically introduces inherent noise on the order of $10^{-5}$ in flux space due to solver limitations, and an MSE of $10^{-8}$ corresponds to differences of $\sqrt{10^{-8}} \approx 10^{-4}$ in the flux unit, already approaching this fundamental precision limitation. For context, even this level of precision (0.01\% in flux) far exceeds what is typically achievable in observational spectroscopy, where signal-to-noise ratios rarely exceed 100 (corresponding to 1\% precision).

\subsection{Caveats and Future Work}

A limitation of our study is considering only the training budget of the spectral emulator when evaluating trade-offs for real surveys. A broader assessment should incorporate both the computational cost of spectral grid calculations and the resources required by the inference stack, the computational process of using trained emulators to make predictions, fit observed spectra, and derive stellar parameters. Expanding the scaling laws derived here to explicitly include grid computation and inference costs could affect optimality and presents interesting technical opportunities for improving the efficiency of future spectral surveys.

Our evaluation focused primarily on how well models can emulate synthetic spectra by calculating the error in normalized flux space. However, this approach is limited to measuring the MSE across entire spectra, while real-world spectroscopic applications are primarily concerned with the precision of inferred stellar parameters and elemental abundances derived from individual spectral lines. For large-scale spectroscopic surveys, where precision requirements are often specified in terms of parameter accuracy rather than spectral MSE, our approach provides a pathway to translate those requirements into concrete computational and data needs. 

We focus on MSE as our primary metric because emulation errors manifest as systematic noise in the spectral fitting process, which, to a first approximation, can be translated into statistical errors on stellar parameters using Fisher matrix calculations. This direct connection between spectral emulation quality and parameter inference precision makes MSE a robust and theoretically justified optimization target \citep{2017ApJ...843...32T,Sandford_2020,2024arXiv240705751R}.

Perhaps the most important limitation is that in this study we use MSE as a proxy for model performance in the training domain only, that is, how well the model performs on synthetic spectra generated using the same methodology as the training data. While optimizing emulation to achieve MSE of $10^{-6}$ (equivalent to approximately 0.1\% in flux) is commendable, such that when dealing with spectra with signal-to-noise ratios of order 100, ensuring the error budget from emulation does not dominate, scaling beyond that point may yield diminishing returns. The critical question for spectral foundational models is domain transfer capability: how well models trained on one domain can generalize to another with minimal additional training.

In the context of spectral emulation, a key challenge is how well models trained on computationally inexpensive 1D-LTE models can transfer to more accurate 3D non-LTE models or to empirical spectra with ground truth labels established through other means \citep[e.g., Gaia Benchmark stars,][]{2024AA...682A.145S}. The original TransformerPayne paper has shown improved domain transfer ability compared to other methods, but this capability remains suboptimal. As we scale up model size following the laws established in this work, understanding how domain transfer ability improves is a critical direction for future research. Similar to how scaling laws in language models have demonstrated emergent abilities like few-shot learning \citep{2020arXiv200514165B, 2024arXiv240217193Z} and ``grokking'' \citep[sudden performance improvements after extended training, e.g.,][]{2022arXiv220102177P}, we aim to investigate whether similar phenomena emerge in spectral models as they scale, potentially enabling more efficient domain adaptation.

\section{Conclusions}
\label{sec:conclusion}

On-going and upcoming spectroscopic surveys produce stellar spectra at unprecedented volumes, demanding computationally efficient techniques for precise analysis. Emulation has become essential in inference pipelines due to its computational speed. It effectively handles the complexity of high-dimensional parameter spaces, overcoming the limitations of traditional interpolation methods. However, earlier emulator models exhibited limitations, with accuracy saturating around a 1\% precision level. Recently, the Transformer-based emulator (TransformerPayne) was introduced to address this limitation by leveraging the inherent scalability and strong inductive bias of Transformer architectures, with the ultimate goal of developing spectral foundational models capable of few-shot learning and domain transfer.

In this work, we empirically investigated how TransformerPayne's emulation accuracy scales with model size, training compute, and the size of the stellar spectra grid. We optimized hyperparameters and employed techniques to stabilize training across all scales, systematically examining improvements in precision.

We established a training methodology based on Maximum Update Parametrization that enables stable and efficient training of large TransformerPayne models. This approach effectively stabilizes training dynamics, allowing consistent scaling of model widths without repeated hyperparameter tuning. Our results demonstrate that once models reach sufficient depth ($N \ge 4$), the primary predictor of model performance is the total parameter count, with considerable flexibility in the exact depth-to-width ratio.

The core result of our empirical investigation is that improvements in emulation quality follow predictable power-law relationships. We have established clear scaling laws for spectral emulation, finding that validation loss scales with dataset size as $\mathcal{L}(D) \propto D^{-1.34}$, with model size as $\mathcal{L}(P) \propto P^{-1.21}$, and with compute as $\mathcal{L}(C) \propto C^{-0.76}$. Along the optimal compute frontier, we found that dataset size should scale as $D \propto C^{0.38}$ and model size as $P \propto C^{0.61}$. These quantitative relationships provide practical guidance for efficiently allocating computational resources to achieve specific emulation precision targets.

The exhibition of these scaling laws demonstrates that Transformer-based models are robust architectures for spectral emulation, with predictable performance improvements as computational resources increase. Our work establishes a foundation for training spectral foundational models without wasteful hyperparameter exploration, offering a clear pathway toward reducing emulation errors to any desired threshold. This capacity supports rapid, reliable spectral interpolation, thereby facilitating more advanced analyses within various spectroscopic surveys.

\section*{Acknowledgments}

This research was undertaken with the assistance of resources and services from the National Computational Infrastructure (NCI), which is supported by the Australian Government. This research used resources of the Oak Ridge Leadership Computing Facility at the Oak Ridge National Laboratory, which is supported by the Office of Science of the U.S. Department of Energy under Contract No. DE-AC05-00OR22725. Y.S.T is supported by the National Science Foundation under Grant No. AST-2406729. This research was supported by grants from NVIDIA Academic Grant Program.  Additionally, we extend our heartfelt thanks to Alex Ji, David Weinberg, Hans-Walter Rix, Jennifer Johnson, Anil Pradhan, Adam Wheeler, Anish Amarsi, Jiri Kubat, Ewa Niemczura, Mark Krumholz, Maria Bergemann, Nicholas Storm, Richard Hoppe and Philip Eitner for invaluable discussions.

\subsection*{Software:}
This project extensively used various open-source Python libraries: numpy \citep{harris2020array}, scipy \citep{2020SciPy-NMeth}, jax \citep{jax2018github}, flax \citep{flax2020github}, optax \citep{deepmind2020jax}, and matplotlib \citep{Hunter2007}. For color maps we mostly used \citet{petroff2024accessiblecolorsequencesdata}.

\bibliography{scaling_laws_transformerpayne}
\bibliographystyle{aasjournal}

\begin{appendix}
\section{Parameter and FLOP Calculations}
\label{appendix:params_and_flops}
In this section, we provide a detailed breakdown of how parameters and FLOPs are computed for the basic building blocks of TransformerPayne. These calculations form the foundation for our scaling law analysis, as they allow us to quantify the computational resources required by models of different sizes. The notation follows that used in the main text.

A linear layer, the fundamental computation unit in neural networks, transforms input vectors from $\mathbb{R}^{d_\text{in}}$ to $\mathbb{R}^{d_\text{out}}$ via matrix multiplication. The number of trainable parameters in such a layer is $ d_\text{in} \times d_\text{out}$. When applied to a single vector $\mathbf{x} \in \mathbb{R}^{d_\text{in}}$, the computational cost can be measured in FLOPs. Counting a multiply plus an add as two FLOPs, the total number of operations is thus $2 \, d_\text{in} \, d_\text{out}$.

For efficiency, neural networks often process multiple inputs simultaneously. When the same linear layer is applied to a matrix $\mathbf{X} \in \mathbb{R}^{L \times d_\text{in}}$ (representing a batch of $L$ inputs), the output becomes $\mathbf{Y} \in \mathbb{R}^{L \times d_\text{out}}$. Though the parameter count remains unchanged, the computational cost scales with the number of inputs of $2 \, L \, d_\text{in} \, d_\text{out}$.

In TransformerPayne, we use RMSNorm for layer normalization but without trainable scale or bias parameters. This simplifies the function computed by RMSNorm on a vector $\mathbf{x} \in \mathbb{R}^d$ to:
\begin{equation}
\text{RMSNorm}(\mathbf{x}) \;=\;
\frac{\mathbf{x}}{\sqrt{\frac{1}{d}\,\sum_{i=1}^{d} x_i^2 \;+\; \epsilon}} \,,
\end{equation}
where $\epsilon$ is a small constant (e.g., $10^{-6}$) for numerical stability. When applied to a sequence of tokens, like $\mathbf{p}_{\text{emb}} \in \mathbb{R}^{t \times d}$, the normalization is applied to each row (token) independently.

Since our implementation of RMSNorm has no learnable parameters, its parameter count is zero. The FLOP count for applying RMSNorm to a $(N_{\text{rows}},\,d)$-matrix involves three main steps: summing the squares of each element (approximately $2d$ operations per row), taking a reciprocal square root once per row, and multiplying each element by the resulting scale factor ($d$ multiplications per row). Combining these operations, the total FLOP count approximates to $3(N_{\text{rows}} \cdot d)$.

The attention mechanism is the core innovation of Transformer architectures. For TransformerPayne, we calculate the computational cost when $Q \in \mathbb{R}^{N_{\text{flux}} \times d}$ and $K,\,V \in \mathbb{R}^{t \times d}$. The attention operation computes:
\[
\text{softmax}\Bigl(\,\frac{QK^\top}{d}\,\Bigr) \;V\,:
\]

This breaks down into three main steps:
\begin{enumerate}
    \item \textbf{Compute} $QK^\top$:
    The multiply-add cost for this matrix multiplication is approximately $2 \times N_{\text{flux}} \times t \times d$ FLOPs.

    \item \textbf{Scale and softmax}:
    Each of the $N_{\text{flux}} \cdot t$ elements is scaled by $1/d$ and then passed through the softmax function along each row. This typically counts as some small constant multiple of $N_{\text{flux}} \cdot t$ (for exponentiation, summation, division, etc.).

    \item \textbf{Multiply by $V$}:
    The attention matrix is multiplied by $V$, with a cost of approximately $2 \times N_{\text{flux}} \times t \times d$ FLOPs.
\end{enumerate}

In multi-head attention, $Q, K, V$ often come from distinct linear projections, and the output may also be projected (the $O$ matrix). We account for these as separate rows in our parameter/FLOP table (e.g., ``Cross-Attn: Q + O'' vs. ``Cross-Attn: K + V''), along with the ``Scores + Context'' step for softmax and multiplication by $V$.

All remaining entries in the main table are derived by combining these base calculations with the appropriate dimensions, multipliers (e.g., the feed-forward layer has $d_{\text{ff}} = 4\,d$) and the correct token counts ($t$ or $N_{\text{flux}}$). We then sum over $N$ layers to arrive at the total parameter and FLOPs counts in the final rows of the table.

These detailed calculations allow us to precisely measure the computational requirements of different model architectures, enabling the systematic investigation of scaling laws presented in the main text.

\section{Scaling TransformerPayne: Additional Materials}
\label{appendix:scaling_tpayne}

This appendix provides supplementary figures that comprehensively illustrate key aspects of TransformerPayne scaling behaviors, expanding upon details briefly discussed in the main text. These visualizations offer deeper insights into the relationships between architectural choices, training dynamics, and model performance that underpin our scaling law analysis.

\begin{center}
    \begin{figure*}[!t]
        \centering
    	\includegraphics[width=1.0\hsize]{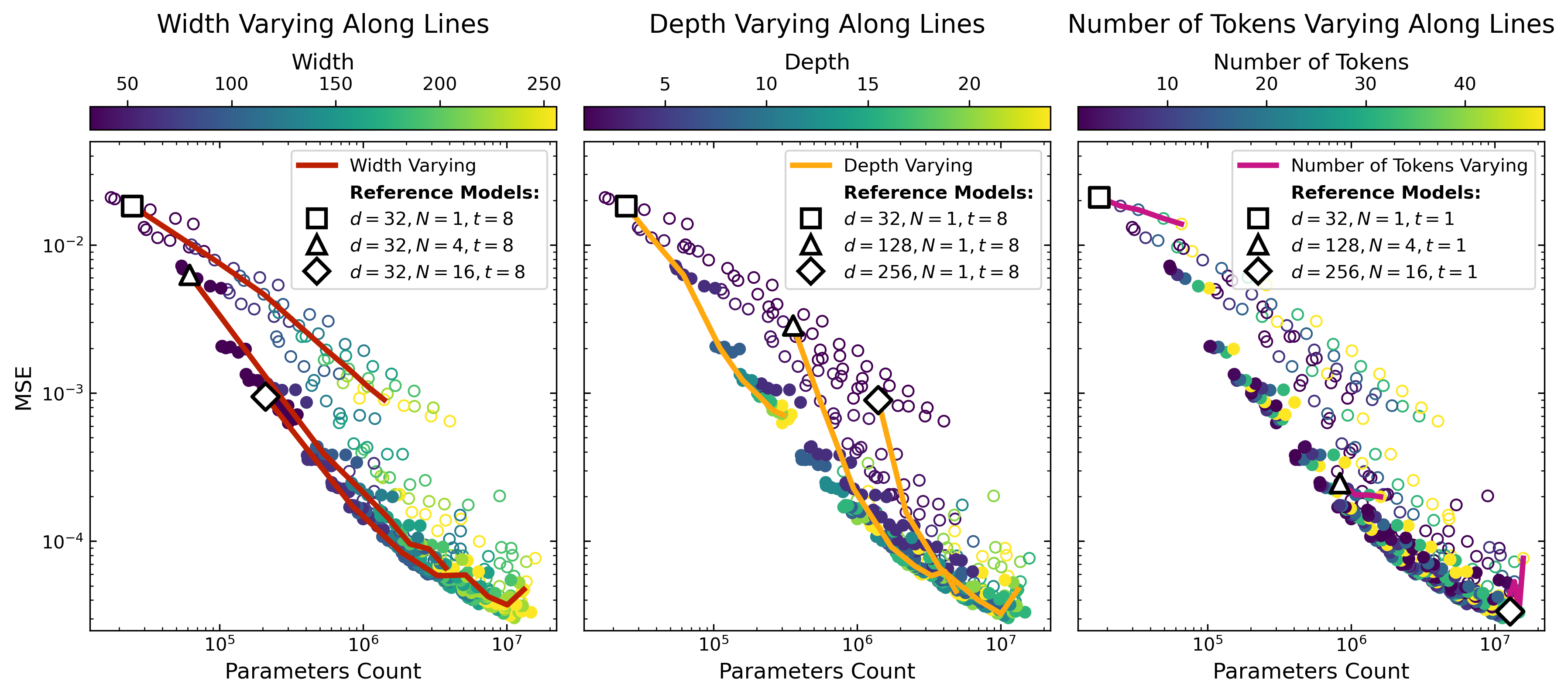}
    	\caption{Validation loss (MSE) as a function of model parameter count, shown across three panels. Each panel highlights how performance varies predominantly with one architectural dimension: model width $d$ (left), depth $N$ (center), and number of tokens $t$ (right). Changes are highlighted using color, and three additional lines are emphasized, each corresponding to variations in only the parameter of interest. Filled markers indicate stable models with sufficient depth ($N \ge 4$) that converged during training, while open markers represent models that were either too shallow ($N < 4$), unstable, or both. The colorbar above each panel shows the corresponding value of the highlighted parameter (width, depth, or tokens). Increasing model width consistently improves performance until instability appears at large widths (at least for largest model). Depth scaling beyond $N \ge 4$ initially offers similar benefits to scaling width but encounters instability earlier compared to width scaling. Varying number of tokens usually yields performance improvements, but with comparatively less benefit than scaling width or depth.
        }
        \label{fig:fig9_scaling_tpayne_shape_detail}
    \end{figure*}
\end{center}

Figure \ref{fig:fig9_scaling_tpayne_shape_detail} provides a more detailed examination of how scaling various architectural parameters affects emulation MSE. While Fig.\,\ref{fig:fig5_scaling_tpayne_shape} in the main text gives an overall view of parameter scaling, this figure breaks down the analysis into three distinct panels, each focusing on one architectural dimension: width (left), depth (center), and token count (right). This decomposition allows us to clearly see the individual contribution of each parameter to model performance.

As discussed in Sec.\,\ref{sec:results_small_models}, width scaling provides consistent improvements with the least risk of instability, depth scaling offers comparable benefits but becomes unstable more quickly, and dedication extra parameters to token count scaling provides the least improvement relative to parameter count. For practical purposes, prioritizing width scaling while maintaining sufficient depth (at least 4 Transformer blocks) offers the most reliable path to improved performance.

\begin{center}
    \begin{figure*}[!t]
        \centering
    	\includegraphics[width=1.0\hsize]{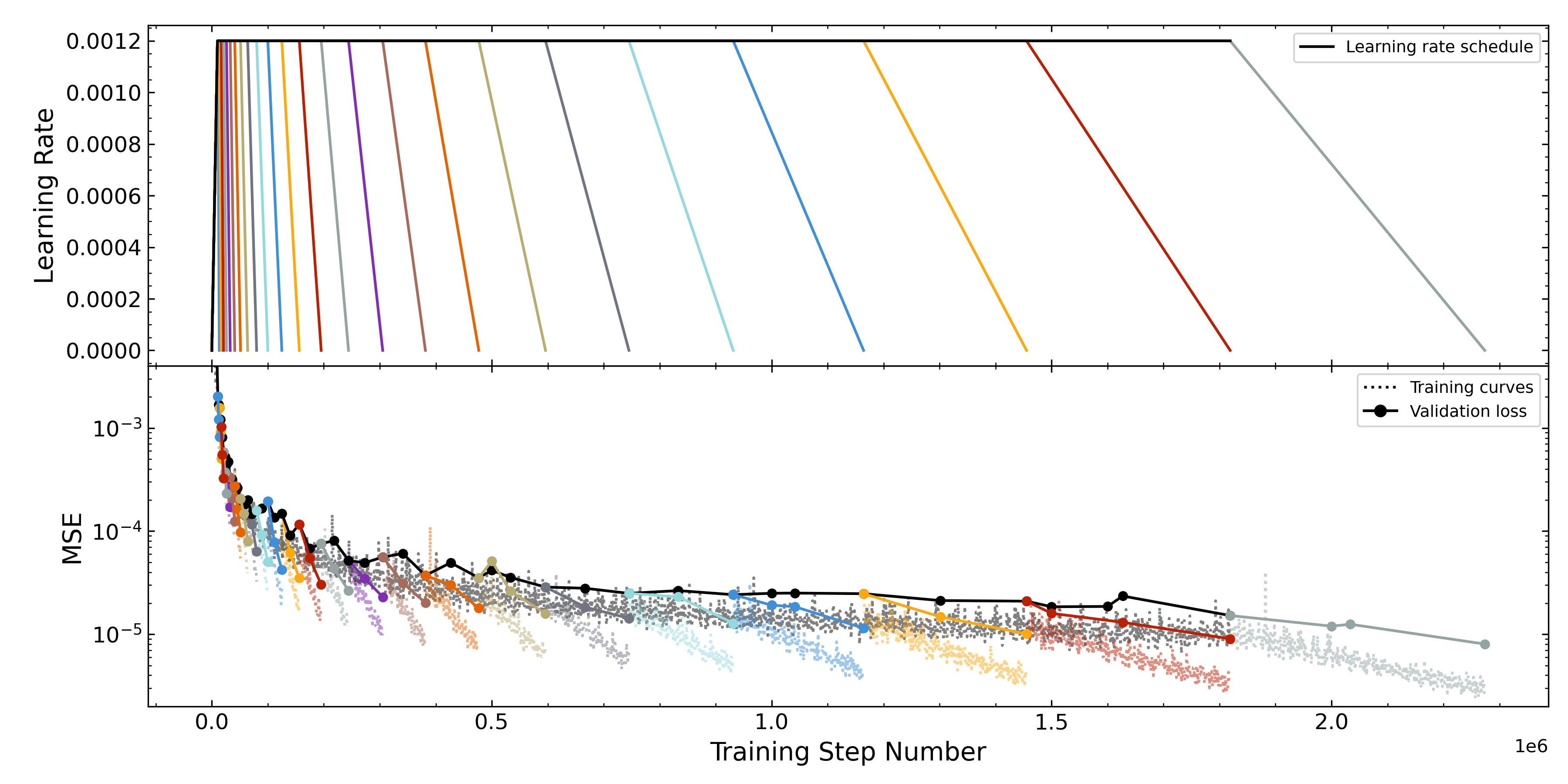}
    	\caption{Visualization of the WSD learning rate schedule (upper panel) and corresponding training and validation loss curves (lower panel) for a TransformerPayne with 4.1M parameters trained on a full dataset of 100,000 spectra. The black line in the upper panel represents the initial warm-up and stable phases, which are shared by all runs. The colored lines show individual cool-down phases, during which each run independently decays the learning rate to zero at distinct restoring and ending steps. In the lower panel, dashed lines represent training loss curves, while solid lines with markers indicate validation loss curves. Colors in the lower panel correspond to the individual cool-down schedules depicted above. This WSD schedule structure was consistently applied to all experiments presented in Sec.\,\ref{sec:results_scaling}. Models at the end of each cool-down phase trace an envelope representing the lowest MSE achievable for each training duration.}
        \label{fig:fig10_wsd_explained}
    \end{figure*}
\end{center}

Figure \ref{fig:fig10_wsd_explained} illustrates the Warmup-Stable-Decay (WSD) learning rate schedule employed in our experiments and its impact on training dynamics. This schedule consists of three phases: an initial warm-up phase where the learning rate increases linearly from zero to its maximum value, a stable phase where the learning rate remains constant, and a final decay phase where it decreases linearly to zero. The upper panel shows how we implemented this schedule, with all runs sharing the same warm-up and stable phases (black line) but having individually customized decay phases (colored lines). The lower panel demonstrates the corresponding training loss (dashed lines) and validation loss (solid lines with markers) for each schedule variant.

This visualization is particularly important for understanding our checkpoint reuse strategy: we ran a single long training process with frequent intermediate checkpoints, then launched multiple shorter decay runs from these checkpoints in parallel. This approach significantly reduced computational costs by reusing the common initial portions of training. The figure clearly shows that the final decay to zero learning rate is crucial for achieving optimal performance, which is why we adopted this schedule for all scaling experiments in Sec.\,\ref{sec:results_scaling}.

\begin{center}
    \begin{figure*}[!t]
        \centering
    	\includegraphics[width=1.0\hsize]{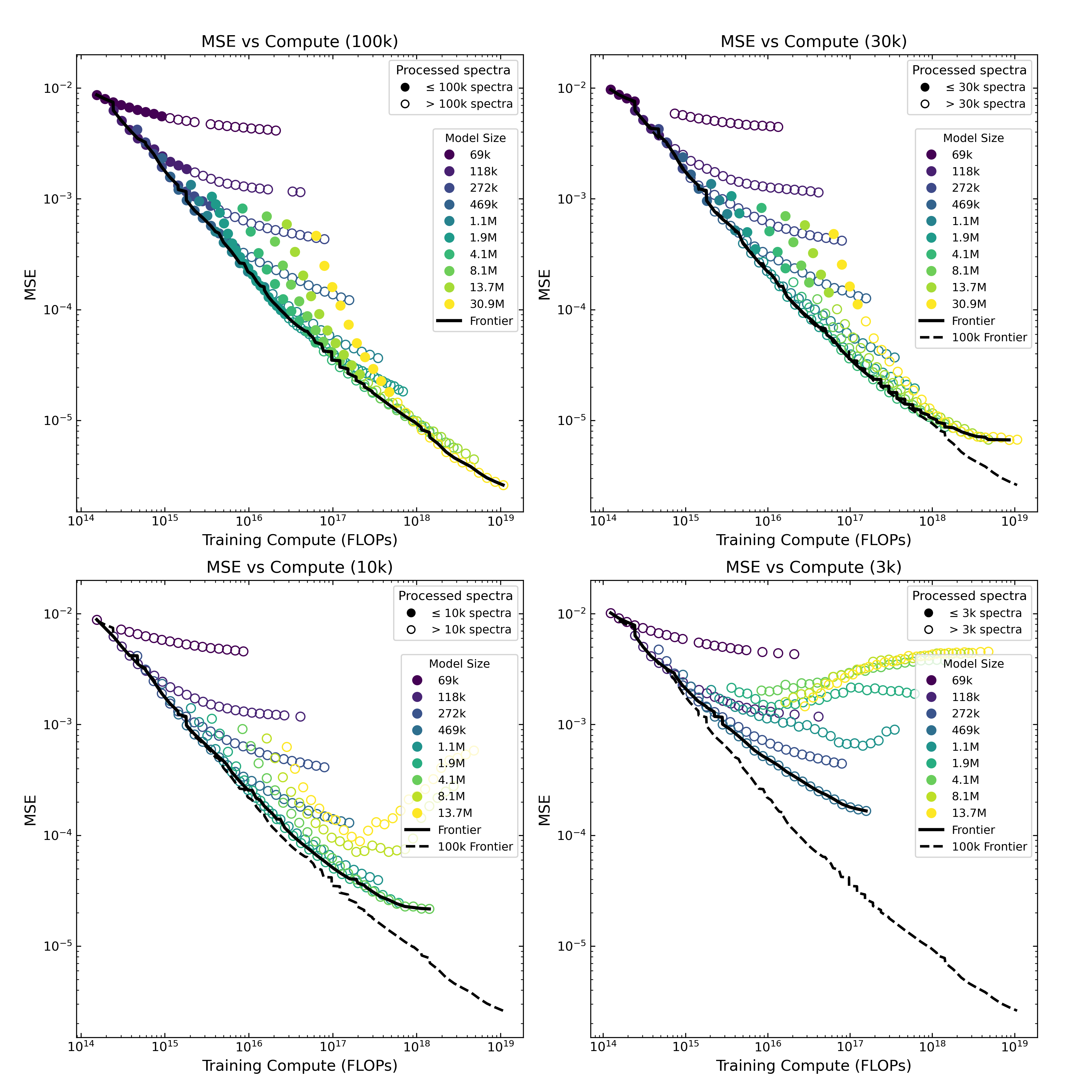}
    	\caption{Scaling experiments illustrating how emulation accuracy changes as the training dataset size decreases. Results are presented for datasets containing 100,000 (top-left panel), 30,000 (top-right), 10,000 (bottom-left), and 3,000 (bottom-right) unique stellar spectra. Each point represents a completed training run, with colors indicating model size (from purple for smallest to yellow for largest models, as shown in the legend). The frontier models, representing the best-performing configurations identified from the dataset of 100,000 spectra, are highlighted with a dashed line across all panels for comparison, allowing direct visualization of how performance degrades with smaller datasets even when maintaining optimal training strategies. Open markers indicate experiments for which training surpassed one epoch (processed more examples than exist in the dataset), while filled markers indicate training within one epoch; notably, all points in the bottom panels correspond to training exceeding one epoch, as smaller datasets are quickly exhausted during training. 
        }
        \label{fig:fig11_raw_scaling_examples}
    \end{figure*}
\end{center}

Figure \ref{fig:fig11_raw_scaling_examples} illustrates how emulation accuracy changes as the training dataset size decreases. Results are presented for datasets containing 100,000 (top-left panel), 30,000 (top-right), 10,000 (bottom-left), and 3,000 (bottom-right) unique stellar spectra. It is important to clarify that term \textit{processed spectra} refers to the total number of training examples seen during training (calculated as batch size $\times$ number of training steps), while the dataset size refers to the number of unique spectra available for training. When the number of processed spectra exceeds the dataset size, the model begins revisiting examples it has already seen.

For the larger datasets (100,000 and 30,000), we observe a mix of filled circles (representing training within one epoch, where each spectrum is seen at most once) and open circles (representing training beyond one epoch, where some spectra are revisited). However, for smaller datasets (3,000 and 10,000), training surpasses one epoch very early, even during the warm-up phase, resulting in only open markers in the corresponding panels.

These plots highlight several critical impacts of limited dataset sizes on scaling behavior. First, while performance frontiers follow similar patterns when sufficient data are available, models trained on smaller datasets increasingly deviate from the initial scaling trends as training progresses, eventually saturating in performance. Second, for smaller datasets, the largest models (shown in yellow/green) exhibit clear signs of overfitting, with MSE values that plateau or even increase with additional training. Third, the best-performing models for smaller datasets typically have intermediate sizes rather than being the largest models available, demonstrating that model capacity should be matched to dataset size.

The early saturation observed in the largest models could likely be mitigated by employing stronger regularization techniques during training, such as increased weight decay, dropout, or early stopping based on validation performance, which is beyond the scope of this study. These results underscore the importance of balancing model size, dataset size, and training duration, particularly when working with limited training data.

\end{appendix}
\end{CJK*}
\end{document}